\documentclass[10pt]{article}

\usepackage{fullpage}
\usepackage{setspace}
\usepackage{parskip}
\usepackage{titlesec}
\usepackage[section]{placeins}
\usepackage{xcolor}
\usepackage{breakcites}
\usepackage{lineno}
\usepackage{hyphenat}
\usepackage{amssymb}
\usepackage{tabularx}
\usepackage{amsmath}
\usepackage{algorithm}
\usepackage{algorithmic} 
\usepackage{subfig}
\PassOptionsToPackage{hyphens}{url}
\usepackage[colorlinks = true,
            linkcolor = blue,
            urlcolor  = blue,
            citecolor = blue,
            anchorcolor = blue]{hyperref}
\usepackage{etoolbox}


\usepackage[authoryear,round]{natbib}

\renewenvironment{abstract}
  {{\bfseries\noindent{\abstractname}\par\nobreak}\footnotesize}
  {\bigskip}

\titlespacing{\section}{0pt}{*3}{*1}
\titlespacing{\subsection}{0pt}{*2}{*0.5}
\titlespacing{\subsubsection}{0pt}{*1.5}{0pt}

\usepackage{authblk}

\usepackage{graphicx}
\usepackage[space]{grffile}
\usepackage{latexsym}
\usepackage{textcomp}
\usepackage{longtable}
\usepackage{tabulary}
\usepackage{booktabs,array,multirow}
\usepackage{amsfonts,amsmath,amssymb}
\providecommand\citet{\cite}
\providecommand\citep{\cite}

\newif\iflatexml\latexmlfalse

\AtBeginDocument{\DeclareGraphicsExtensions{.pdf,.PDF,.eps,.EPS,.png,.PNG,.tif,.TIF,.jpg,.JPG,.jpeg,.JPEG}}

\usepackage[utf8]{inputenc}
\usepackage[english]{babel}

\usepackage{float}

\begin{document}

\title{Analytical Optimized Traffic Flow Recovery for Large-scale Urban Transportation Network}

% \author[1]{Sicheng Fu}%
% \affil[1]{Affiliation not available}%

\author[1]{Sicheng Fu}
\author[1,*]{Haotian Shi}
\author[1]{Shixiao Liang}
\author[2]{Xin Wang}
\author[1]{Bin Ran}

\affil[1]{Department of Civil and Environmental Engineering, University of Wisconsin-Madison, Madison,53706, WI, USA}

\affil[2]{Department of Industrial and Systems Engineering, University of Wisconsin-Madison, Madison,53706, WI, USA}

\vspace{-1em}

  % \date{June 21, 2024}

\begingroup
\let\center\flushleft
\let\endcenter\endflushleft
\maketitle
\endgroup

\renewcommand{\thefootnote}{\fnsymbol{footnote}} 
\footnotetext[1]{Corresponding author: Haotian Shi (hshi84@wisc.edu)}

\selectlanguage{english}
\begin{abstract}
The implementation of intelligent transportation systems (ITS) has enhanced data collection in urban transportation through advanced traffic sensing devices. However, the high costs associated with installation and maintenance result in sparse traffic data coverage. To obtain complete, accurate, and high-resolution network-wide traffic flow data, this study introduces the Analytical Optimized Recovery (AOR) approach that leverages abundant GPS speed data alongside sparse flow data to estimate traffic flow in large-scale urban networks. The method formulates a constrained optimization framework that utilizes a quadratic objective function with $l_{2}$ norm regularization terms to address the traffic flow recovery problem effectively and incorporates a Lagrangian relaxation technique to maintain non-negativity constraints. The effectiveness of this approach was validated in a large urban network in Shenzhen’s Futian District using the Simulation of Urban MObility (SUMO) platform. Analytical results indicate that the method achieves low estimation errors, affirming its suitability for comprehensive traffic analysis in urban settings with limited sensor deployment.
\end{abstract}%

\sloppy

\section{Introduction}
\label{sec:Introduction}
The acquisition of detailed, accurate, and high-resolution traffic flow data has become a fundamental requirement in modern urban transportation systems, eliciting interest from governmental agencies and commercial sectors. The value of such data is manifold: it enables enhanced planning and management of urban traffic \citep{di2016network}, facilitates infrastructure development and construction initiatives by government bodies \citep{unterluggauer2022electric}, empowers commuters with the information necessary to make well-informed travel decisions \citep{chao2014intelligent}, and supports the formulation of operational decisions and policies aimed at augmenting the safety, efficiency, and environmental sustainability of intelligent transportation systems (ITS) \citep{iordanidou2014feedback, zhu2022research}.

To obtain real-time traffic flow data, ITS in urban networks utilize advanced technologies and devices such as vehicle loop detectors \citep{zhang2003development}, traffic surveillance cameras, \citep{aasen2018quantitative}, and Roadside Units (RSUs) equipped with V2X (Vehicle-to-everything) technology \citep{khan2017data}. While these devices are effective at monitoring traffic flow data, their widespread deployment and implementation in diverse environments pose significant economic and logistical challenges. This results in an inability to obtain accurate, city-level traffic flow data on a large scale \citep{fabris2024efficient}. Specifically, inductive loop detectors necessitate invasive roadwork for their installation and sustain high maintenance costs, particularly in high-traffic areas. Cameras, while offering visual traffic insights, are susceptible to occlusions and can suffer performance degradation due to adverse weather or insufficient lighting. RSUs, which furnish localized data, can be limited by their range, leading to incomplete data sets that may not reflect the entire traffic landscape due to potential communication losses. These factors contribute to financial and technological barriers that limit the widespread deployment of sensing devices. Consequently, this leads to the collection of inaccurate and incomplete traffic data, hindering the ability of ITS to gather comprehensive and high-resolution traffic flow information across urban networks. These shortcomings present substantial challenges for traffic management, route planning, and decision-making processes across stakeholders, including enterprises, government agencies, individual vehicles, and cloud-based systems \citep{bibri2021data}.

To address the issue, the widespread availability of GPS data provides a viable alternative for traffic flow recovery \citep{necula2014dynamic, zhou2013traffic}, as there is often a strong correlation between traffic flow and link-level speed data \citep{zhou2013traffic, essien2021deep}. This link-level speed data, captured by GPS, is more economically feasible and more extensively available, offering the potential to fill the gaps in traffic flow data collection. When analyzed, GPS data could potentially reduce the reliance on expansive physical sensor networks. Furthermore, the growing prevalence of Connected and Autonomous Vehicles (CAVs) and the wide use of GPS-enabled devices have created a rich repository of link-level speed data. This data can be leveraged to deduce comprehensive traffic flows across large-scale urban areas. Therefore, it is critical to effectively utilize multi-source traffic data by combining GPS data with sparsely observed traffic flow data to enhance the precision and reliability of traffic flow recovery within large-scale urban traffic networks.

To leverage the potential of GPS data for enhanced traffic flow recovery, it is crucial to integrate this data with traffic flow recovery methods.  Traffic flow recovery refers to the process of reconstructing traffic flow patterns across transportation networks using sparse or incomplete sources. Conventional research on traffic flow recovery primarily employs three methodologies to reconstruct flow patterns within transportation networks: vision-based techniques using image data, data-driven methods utilizing machine learning, and traditional model-based approaches grounded in optimization techniques. These methods are thoroughly reviewed in the literature review section, referenced in section \ref{sec:litreview}. While these studies have advanced traffic monitoring and management within urban transportation networks, several critical gaps exist. Firstly, the complex characteristics of urban traffic networks, with their heterogeneous traffic flows across multiple road types, present challenges for traffic flow recovery. Most studies simplify their focus to a single road type, such as highways, which is insufficient to accurately depict the dynamic flow progression of the entire large-scale urban traffic network. Secondly, many methods require extensive historical data for model training. Consistently and reliably collecting such data in urban environments is challenging, and the data often centers on major roads and typical traffic flow patterns, lacking comprehensive coverage. Lastly, the scalability of these methods to city-wide applications raises concerns due to the prohibitive computational resource requirements. Certain techniques struggle with simulations under these constraints, hindering their ability to predict future traffic conditions or manage traffic effectively in real time.

Addressing these gaps, this paper introduces a novel framework: an Analytical Optimized Recovery (AOR). This framework aims to recover accurate and high-resolution link flow in large-scale urban traffic networks, utilizing available link-level speed data acquired from GPS device-equipped vehicles and sparsely observed traffic flow data from road segments with traffic sensing devices. The relationship between traffic speed and flow is established through a dynamic traffic assignment model, represented as a system of equations correlating the demand and flow on each road segment \citep{ma2020estimating}. Due to the sparse nature of traffic flow observations, these equations yield multiple solutions. Additional constraints are imposed in each framework to refine the recovery. The AOR framework employs a constrained optimization framework with quadratic objective function and $l_{2}$ norms, solving the problem through a linear analytical formulation. This enables the application of Stochastic Gradient Descent (SGD) for hyperparameter tuning, enhancing the model's performance. Moreover, the application of the Lagrangian Relaxation (LR) method ensures the maintenance of non-negativity constraints. 

Through these methods within the framework, this paper provides a robust and interpretable recovery of traffic flow in urban networks, overcoming the limitations of current methods and computational challenges. To evaluate the performance of the proposed models, a detailed traffic network simulation platform using the Simulation of Urban MObility (SUMO) has been developed. This platform, based on authentic traffic conditions, calibrated with real data, and encompassing various road types within a city-level network, generates a large dataset for experimental purposes. By acting as a digital twin platform, it facilitates the assessment of the model's applicability to real-world scenarios.

The contributions of this work are summarized as follows:
\begin{itemize}

\item Develop a framework based on constrained optimization that utilizes a quadratic objective function with $l_{2}$ norm regularization terms for the analytical recovery of dynamic high-resolution link flow, using traffic speed data and a limited set of observed link flows.

%\item Design an efficient parallel computation algorithm to represent the dynamic traffic flow propagation in the large-scale urban road network

\item Utilize the refined Lagrangian relaxation method to effectively handle non-negativity constraints and refine hyperparameters, thereby boosting the accuracy of traffic flow estimations.

\item Build a large-scale digital twin of an urban road network using the simulation platform SUMO for data generation and experimental analysis. The proposed methodology is validated based on the platform.

\end{itemize}

The structure of the paper is outlined as follows: Section 2 provides a comprehensive literature review, surveying existing research relevant to traffic flow recovery. Section 3 details the analytical recovery optimization framework, discussing its methodology. Section 4 describes the associated experiments conducted. Finally, Section 5 concludes the paper and outlines future research directions.

\section{Literature Review}
\label{sec:litreview}

This literature review section aims to explore the multifaceted approaches developed to estimate and recover traffic flows, focusing on advancements in vision-based approaches, machine learning-based methods, and optimization frameworks in flow recovery problems. 
%as shown in Table \ref{table:research_comparison}.

\subsection{Computer Vision Based}
The prevailing method for gathering traffic-related data is the deployment of traffic surveillance cameras, which have attracted significant research interest due to computer vision technologies \citep{buch2011review,manguri2023review}.
Early methodologies utilizing traditional computer vision techniques for vehicle counting and tracking have been established during freeway congestion \citep{beymer1997real} and transition from shadow to lighting \citep{coifman1998real}. In the early 20th century, there are notable work by \cite{khoshabeh2007multi}, who employed an omnidirectional camera alongside a pan-tilt-zoom (PTZ) camera to analyze traffic flows. Additionally, \cite{bas2007automatic} introduced a method that adjusts bounding box sizes for vehicle detection and tracking based on the vehicle's estimated distance from the camera, factoring in scene-camera geometry. \cite{xia2016towards} integrated the expectation-maximization algorithm with the Gaussian mixture model for better vehicle segmentation to achieve counting accuracy. As machine learning technologies have evolved, these vision-based methodologies have been expanded and refined in further research endeavors. For example, \cite{ide2016city} introduced an unsupervised learning model to determine vehicle numbers from low-quality images and infer network-wide flows using Markov formulation.  Additionally, some researchers considered designing clustering algorithms \citep{lin2021vehicle} and using edge computing to enhance computer vision in traffic flow recovery based on video data \citep{liu2021smart}

Furthermore, the advancement of GPU technology has significantly enhanced the capabilities of vision-based traffic estimation through the application of complex deep learning architectures, such as Convolutional Neural Networks (CNNs) and Recurrent Neural Networks (RNNs). For instance, \cite{wei2019city} developed a real-time system that leverages deep learning pipelines to track vehicles and estimate traffic flow using traffic cameras that capture low frame rates and low-resolution videos. Similarly, \cite{fedorov2019traffic} employed a neural network model combined with a vehicle tracker to count vehicles and analyze their directional movements within urban networks. \cite{tang2022spatiotemporal} introduced the STGGAT model, a sophisticated deep-learning approach that predicts traffic flow by analyzing license plate data to gather traffic metrics. Additionally, \cite{qi2019automated} implemented object detection models using the Single Shot MultiBox Detector (SSD) framework to enhance traffic monitoring. Despite these advancements, the deployment of surveillance cameras remains insufficiently dense to cover most road segments within urban transportation networks comprehensively. Consequently, relying solely on limited camera data to extrapolate traffic conditions for unmonitored road segments often leads to inaccuracies. 

\subsection{Data Driven Based}
Moreover, some other existing works have sought to estimate traffic patterns by combining surveillance cameras and GPS devices to observe information, especially relying on the machine learning and data driven based approach. In the earlier work, traffic flow estimation and prediction may use some computational methods. \cite{dharia2003neural} introduced a neural network model for forecasting freeway link travel time using a counterpropagation network. \cite{jiang2004wavelet} presented a wavelet packet-autocorrelation function for traffic flow pattern analysis. Then \cite{jiang2005dynamic} further developed a dynamic wavelet neural network model to provide accurate and timely forecasts. 

Recent advancements have seen a surge in the exploration of supervised learning approaches by researchers. \cite{meng2017city} introduced a spatio-temporal semi-supervised learning model that leverages data from loop detectors and taxi trajectories to accurately estimate traffic volumes for individual roads. Similarly, \cite{li2021multi} assessed the efficacy of the Gaussian Process Regressor (GPR) in reconstructing traffic flows from estimated travel times, utilizing crowd-sourced data for this purpose. Besides, neural networks have emerged as a pivotal tool in the recovery and prediction of comprehensive traffic flow information. \cite{yi2019citytraffic} proposed a neural memorization and generalization approach to infer the missing traffic speed data derived from the taxi GPS trajectory and volume data derived from camera records. \cite{zhang2020network}
introduced a network-wide traffic flow estimation approach that integrates spatial affinity correlations from crowdsourced floating car data and temporal continuity characteristics within a geometric matrix factorization model to complete missing traffic volume. In addition, a graph neural network (GNN) \citep{lei2022modeling}, generative adversarial network (GAN) \citep{li2018urban} and deep neural network (DNN) \citep{liu2021deeptsp} methods were provided to model the traffic state including position and speed. These pioneering works demonstrate that integrating additional data sources, such as GPS speed has great potential for enhancing the accuracy of large-scale traffic flow estimation. In addition, \cite{kashyap2022traffic} reviewed deep learning techniques that are utilized in traffic flow prediction tasks. 

However, these methodologies predominantly adopt a data-driven perspective, inferring traffic conditions based on spatial and temporal proximity. This approach overlooks critical dynamic traffic flow propagation factors, which play a pivotal role in shaping real-world driver behaviors. The omission of dynamic flow propagation leads to an insufficient comprehension of the route choice decision-making process and overall network performance, especially in a congested network, or considering the variability of driver behavior. Moreover, such model-free strategies necessitate extensive historical data for training, a resource often challenging to procure at the scale of urban environments. Additionally, the complexity inherent in the design of network structures presents significant computational hurdles, typically resulting in prolonged training and validation periods. 

\subsection{Optimization Framework Based}
In addressing the limitations of data-driven methodologies in capturing the intricacies of dynamic traffic flow and driver behavior, traffic assignment models offer a robust alternative \citep{yang2001simultaneous,lu2013dynamic,fan2020dynamically}. From the early foundational work on the User Equilibrium (UE) model\citep{wardrop1952road}, research has evolved to incorporate this concept into traffic flow estimation using a bi-level optimization framework \citep{nguyen1977estimating, leblanc1982selection, yang1992estimation, florian1995coordinate}. This progression led to the development of the Stochastic User Equilibrium (SUE) model, which enhances accuracy by integrating logit-based route-choice probabilities to more precisely estimate travel costs \citep{dial1971probabilistic,daganzo1977stochastic,fisk1980some}. Such advancements in modeling have been instrumental for a detailed analysis of congestion impacts and have contributed significantly to the development of dynamic traffic assignment (DTA) models. These models acknowledge the variability in traffic patterns over time, providing a more realistic representation of traffic dynamics \citep{bell1997stochastic,liu2023stochastic,li2023strategy,daganzo1977some,fisk1977note,daskin1985urban}.

Besides these sophisticated models, the state-of-the-art matrix completion has also advanced traffic flow recovery by addressing data sparsity and sensor limitations \citep{ran2016tensor,yu2020urban,zhang2020network}. Recent developments in matrix completion techniques have leveraged robust statistical methods to predict missing traffic data with greater accuracy \citep{li2022nonlinear}. Innovations such as high-dimensional matrix factorization and deep learning-based tensor completion have enabled the handling of multi-way data arrays, reflecting complex interactions across different dimensions such as time, space, and vehicle type \citep{li2022st}.

However, optimization-based approaches, particularly those involving bi-level programming and dynamic network loading, encounter significant computational challenges that limit their scalability to large road networks. Consequently, many studies focus only on typical highways for traffic flow estimation, which does not adequately represent the heterogeneity of road types across the entire urban network. Moreover, the insufficiency of computational resources often impedes these algorithms from effectively generating virtual data through simulations, thereby obstructing the progress of experimental validations.

This discussion sets the stage for a summarized presentation of the various traffic flow recovery methodologies and their associated gaps. The evolving landscape of traffic flow modeling underscores the need for approaches that combine robust theoretical frameworks with practical applicability to effectively manage urban traffic systems.

\section{Methodology}
This section first describes the problem addressed in this work. Next, it presents the process of constructing a dynamic assignment matrix. Following this, the AOR approach, which integrates a dynamic assignment matrix with an analytical formulation, is provided. Finally, the implementation of optimization techniques, including SGD and LR, is discussed.

\subsection{Problem Statement}
\begin{figure}
    \centering
    \includegraphics[width=0.75\paperwidth]{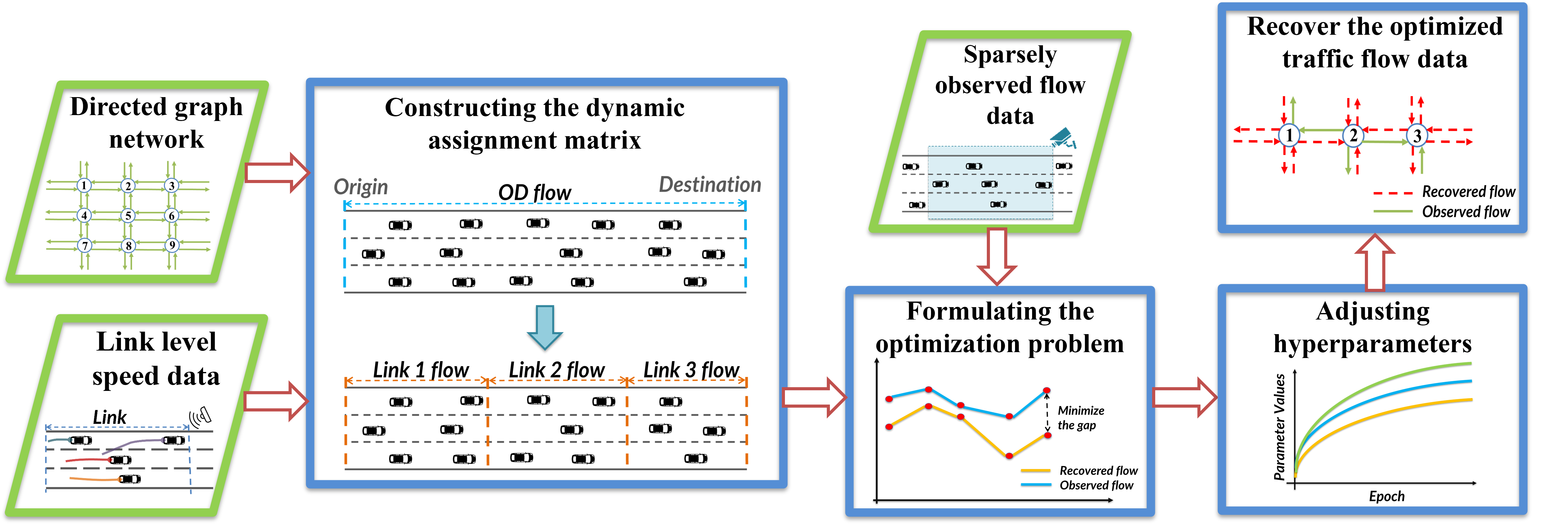}
    \caption{Flowchart of the proposed Analytical Optimized Recovery (AOR) approach}
    \label{fig:flowchart}
\end{figure}

The primary issue addressed in this study is the recovery of link flow in large-scale urban transportation networks using GPS speed data and sparsely observed flow data. Establishing a mathematical relationship between link-level speed data and traffic flow is crucial for accurate estimation. Dynamic traffic assignment models are typically used to distribute origin-destination (OD) flows to network links, which then allows for the derivation of travel times for each link. However, in this study, the availability of GPS data provides direct speed measurements for each link, enabling the direct determination of link travel times. Thus, the objective is to utilize these known travel times in a directed graph traffic network, along with a designed dynamic assignment matrix, to reverse infer the flow information across the urban network. To achieve this, the AOR approach is proposed, which is illustrated in Figure \ref{fig:flowchart}.

In this approach, the directed graph of traffic network information and GPS-captured link-level speed data is used to construct the dynamic assignment matrix, crucial for mapping the interactions between the OD flow and link flow. Once the assignment matrix is generated, it informs an optimization formulation that employs a quadratic objective function and an $l_{2}$ norm to effectively estimate link flows from sparsely observed traffic data. Subsequently, the quadratic objective function is transformed into a linear analytical formulation. This transformation simplifies the model, establishing a more direct and intuitive relationship between the observed and estimated traffic flows, thereby enhancing the model's flexibility and solvability. To refine the model's accuracy and ensure the maintenance of non-negativity in the flow recovery, optimization techniques such as SGD and LR are applied. These methods are instrumental in fine-tuning the hyperparameters and optimizing the recovery performance of the traffic flow model.

Therefore, this work effectively integrates a dynamic assignment model with a quadratic objective function framework, ensuring accurate and efficient traffic flow recovery in large-scale urban transportation networks. The subsequent section will detail the process of constructing the dynamic assignment matrix.

\subsection{Notations}
Refer to Table \ref{tab:notation} for a list of notations that are used in constructing a dynamic assignment matrix.

\begin{table*}[ht]
\caption{List of Notations}
\centering
\begin{tabularx}{\textwidth}{l X}
\hline \textbf{Variables} & \textbf{Description} \\
\hline 
$G$ & Urban transportation network \\
$N$ & Node set \\
$L$ & Link set \\
$(n_{o},n_{d})$ & Origin node to destination node pair \\
$K_{n_{o}n_{d}}$ & Path set for OD pair $(n_{o},n_{d})$ \\
$T$ & Planning horizon \\
$\delta_{l}^{k}$ & Binary indicator that path $k$ includes link $l$ or not \\
$f_{n_{o},n_{d}}^{k,t_{o}}$ & Flow rate on the $k$th path for the OD pair $n_{o},n_{d}$ departing at time $t_{o}$ \\
$x_{l}^{t}$ & Traffic flow at link $l$ at time $t$ \\
$q_{n_{o}n_{d}}^{t_{o}}$ & Travel demand between $n_{o}$ and $n_{d}$ departing at time $t_{o}$ \\
$d_{l}$ & Length for link $l$ \\
$v_{l}(t)$ & Travel speed of link $l$ at time $t$ \\
$\tau_{n_{o},n_{d}}^{k}(t_{o})$ & Travel time from node $n_{o}$ to node $n_{d}$ along path $k$, starting from $t_{o}$ \\
$\rho_{n_{o},n_{d}}^{k,l}(t_{o},t)$ & Fraction of the total flow departing from $n_{o}$ at time $t_{o}$ that would be present on link $l$ at time $t$ \\
$\theta_{n_{o}n_{d}}^{k,t_{o}}$ & Fraction of the overall OD demand $q_{n_{o},n_{d}}^{t_{o}}$ for travel between $n_{o}$ and $n_{d}$ departing at time $t_{o}$ that selects path $k$ \\
\hline
\end{tabularx}
\label{tab:notation}
\end{table*}

\subsection{Dynamic Assignment Matrix} \label{sec:assignment_mat}
To establish an estimation framework that correlates link speed with link flow, it is essential to articulate the relationship between link flow with path flow within a transportation network, denoted by $G=(N,L)$. Here, $L$ is the link set, representing road segments, while $N$ is the node set, signifying the junctures at which road segments intersect, such as intersections. In the static assignment model, let $f_{n_{o},n_{d}}^{k}$ denote the flow rate on the $k$-th path from its origin node $n_{0}\in N$ to its destination node $n_{d} \in N$, and let $x_{l}$ represent the traffic flow on link $l \in L$. The link flow can be construed as the collective sum of the flows on all paths that traverse it, as encapsulated by the Equation \ref{eq1}:

\begin{equation} \label{eq1}
x_{l}=\mathop{\sum_{n_{o}\in N}\sum_{n_{d}\in N}\sum_{k\in K_{n_{o},n_{d}}}\delta_{l}^{k}f_{n_{o},n_{d}}^{k}}
\end{equation}

Where $K_{n_{o}n_{d}}$is the path set for OD pair $n_{o}, n_{d}$, and $\delta_{l}^{k}$ is the binary indicator that 1 if path $k$ includes link $l$ and 0 otherwise. To address the dynamics of link flow evolution and incorporate speed considerations, the static equation must be extended into a temporal dimension, considering traffic conditions over a specified planning horizon $T$. The dynamic formulation of link flow is presented in the equation \ref{eq2}:

\begin{equation} \label{eq2}
x_{l}^{t}=\sum_{t_{o}\in T}\sum_{n_{o}\in N}\sum_{n_{d}\in N}\sum_{k\in K_{n_{o}n_{d}}}\rho_{n_{o},n_{d}}^{k,l}(t_{o},t)f_{n_{o},n_{d}}^{k,t_{o}}
\end{equation}

In the dynamic context, $x_{l}^{t}$ is the traffic flow at link $l \in L$ at time $t \in T$, $f_{n_{o},n_{d}}^{k,t_{o}}$ is the flow rate on the $k$th path for the OD pair $n_{o},n_{d}$ departing at time $t_{o}$, and $\rho_{n_{o},n_{d}}^{k,l}(t_{o},t)$ is a weight function that quantifies the proportion of flow departing from $n_{0}$ at time $t_{o}$ to $n_{d}$ along path $k$, measured at link $l$ at time $t$. The weight function reflects the link-level flow progression from time $t_{0}$ to $t$, capturing the dynamic behavior of traffic flow.

To obtain the time-varying feature of $\rho_{n_{o},n_{d}}^{k,l}(t_{o},t)$, it is necessary to possess comprehensive information regarding travel times throughout the transportation network. Leveraging GPS technology allows for the real-time capture of the travel speed of each link $l$ at time $t$, denoted by $v_{l}(t)$. This collected speed information is crucial for estimating the time vehicles take to travel across each link. In this analysis, we assume vehicles are distributed evenly across each link, and the rate at which vehicles enter a link is constant over time. The assumption is predicated on the difficulty of calibrating or learning the dynamics of within-link shockwaves without access to detailed trajectory data, particularly within the context of a large-scale network. Indeed, empirical evidence suggests that modeling link-level flow progression is both realistic and yields stable and efficient outcomes. Thus, under steady-state conditions, the travel time $\tau_{l}(t)$ for a link $l$ of length $d_{l}$ is calculated as Equation \ref{eq3}

\begin{equation} \label{eq3}
\tau_{l}(t)=\frac{d_{l}}{v_{l}(t)}
\end{equation}

Given this, the travel time from origin $n_{0}$ along path $k$ to pass link $l$ can be obtained by summing the travel times for each individual link $l_{i}$ that compose the path up to link $l$, as represented by Equation \ref{eq4} 

\begin{equation} \label{eq4}
\tau_{n_{o},n_{d}}^{k,l}(t_{o})=\sum_{l_{i}\in k}\tau_{l_{i}}(t_{l_{i}})
\end{equation}

Where $l_{i}$ denotes the link sequence connecting $n_{o}$ till link l along path $k$ and $t_{l_{i}}$ indicates the corresponding starting time the vehicle arrives at link $l_{i}$. From this foundation, the weight function $\rho_{n_{o},n_{d}}^{k,l}(t_{o},t)$ encapsulates the fraction of the total flow departing from $n_{o}$ at time $t_{0}$ that would be present on link $l$ at time $t$, which defined as Equation \ref{eq5}:

\begin{equation} \label{eq5}
\rho_{n_{o},n_{d}}^{k,l}(t_{o},t)=\frac{\left|\left[\tau_{n_{o},n_{d}}^{k,l}(t_{o}),\tau_{n_{o},n_{d}}^{k,l}(t_{o}+\Delta t)\right]\cap[t,t+\Delta t]\right|}{\tau_{n_{o},n_{d}}^{k,l}(t_{o}+\Delta t)-\tau_{n_{o},n_{d}}^{k,l}(t_{o})}
\end{equation}

Here, $\Delta t$ is the temporal resolution of the sampling. The interval $\left[\tau_{n_{o},n_{d}}^{k,l}(t_{o}),\tau_{n_{o},n_{d}}^{k,l}(t_{o}+\Delta t)\right]$ indicates the expected arrival time at link $l$ for vehicles that depart within the time window $[t_{o},t_{o}+\Delta t]$. The percentage of this interval overlaps with $[t,t+\Delta t]$ indicates the time period during which these vehicles are observed at link $l$ near time $t$. The denominator normalizes the overlap period by the entire duration spanned by the arrival time window. This normalization process effectively translates the observed overlap into a quantifiable metric, representing the proportion of the traffic flow on link $l$ at time $t$. This methodology is instrumental in determining the dynamic characteristics of traffic flow across the network, providing insights into how traffic evolves and disperses over time. 

From $n_{o}$ to $n_{d}$, there can be multiple paths that vehicles can take. Travelers choose among these available paths according to a path choice function, denoted as $\theta_{n_{o}n_{d}}^{k,t_{o}}$. The function determines the fraction of the overall OD demand $q_{n_{o}n_{d}}^{t_{o}}$ for travel between $n_{o}$ and $n_{d}$ departing at time $t_{o}$ that selects path $k$. The distribution of OD demand to path flow is expressed by the following Equation \ref{eq6}:

\begin{equation} \label{eq6}
f_{n_{o},n_{d}}^{k,t_{o}}=\theta_{n_{o}n_{d}}^{k,t_{o}}q_{n_{o}n_{d}}^{t_{o}}
\end{equation}

A typical way to determine $\theta_{n_{o}n_{d}}^{k,t_{o}}$ is to use the logit-based path choice model. This model posits that the likelihood of selecting a particular path is exponentially related to the travel time on that path, encapsulated by the Equation \ref{eq7}: 

\begin{equation} \label{eq7}
\theta_{n_{o}n_{d}}^{k,t_{o}}=\frac{\exp(-\alpha\tau_{n_{o},n_{d}}^{k}(t_{o}))}{\sum_{k\in K}\exp(-\alpha\tau_{n_{o},n_{d}}^{k}(t_{o}))}
\end{equation}

Where $\tau_{n_{o},n_{d}}^{k}(t_{o})$ indicates the travel time from node $n_{o}$ to node $n_{d}$ along path $k$, starting from $t_{o}$, while the parameter $\alpha$ that reflects the sensitivity of travelers to travel time differences among paths. A larger value of $\alpha$ implies a stronger preference for shorter travel times and is typically assigned a default value of 0.01.

Combining Equation \ref{eq2} and Equation \ref{eq6}, the dynamic assignment model formalizes the relationship between link flows and OD demand as Equation \ref{eq8}: 

\begin{equation} \label{eq8}
x_{l}^{t}=\sum_{t_{o}\in T}\sum_{n_{o}\in N}\sum_{n_{d}\in N}\sum_{k\in K_{n_{o},n_{d}}}\rho_{n_{o}n_{d}}^{k,l}(t_{o},t)\theta_{n_{o}n_{d}}^{k,t_{o}}q_{n_{o}n_{d}}^{t_{o}}
\end{equation}

The objective of the dynamic assignment model is to express link flow as a function of the path flows that emanate from all OD pairs and transit the link at a specified time. With help of collected time-varying speed data, $\rho_{n_{o}n_{d}}^{k,l}(t_{o},t)$ and $\theta_{n_{o}n_{d}}^{k,t_{o}}$ can be calibrated and seen them as known parameters. Thus, we can rewrite the Equation \ref{eq8} as its matrix form,

\begin{equation} \label{eq9}
\mathbf{x}=\boldsymbol{\rho}\boldsymbol{\theta}\mathbf{q}
\end{equation}

where $\mathbf{x}\in R^{|L||T|\times1}$, $\boldsymbol{\rho}\in\mathbb{R}^{|L||T|\times|N|^{2}|K||T|}$, $\boldsymbol{\theta}\in\mathbb{R}^{|N|^{2}|K||T|\times|N|^{2}|T|}$, and $\mathbf{q}\in\mathbb{R}^{|N|^{2}|T|\times1}$. Here, $\rho$ is defined as the assignment weight matrix, $\theta$ is defined as the path choice matrix. The vector $\mathbf{x}$ represents the link flow vector we aim to estimate, while $\mathbf{q}$ is the vector representing the estimated OD demand. We further denote $\mathbf{A}$ as the assignment matrix, the entries of $\mathbf{A}$ can be computed as in equation,

\begin{equation} \label{eq10}
\mathbf{A}=\boldsymbol{\rho}\boldsymbol{\theta}
\end{equation}

where $\mathbf{A}\in\mathbb{R}^{|L||T|\times|N|^{2}|T|}$. The assignment matrix $\mathbf{A}$ can be conceptualized as the product of the assignment weight matrix $\boldsymbol{\rho}$ and the path choice matrix $\boldsymbol{\theta}$, effectively integrating the influences of both route preference and traffic distribution across the network. The sparse matrix $\mathbf{A}$ effectively maps OD demand directly to the link flows and is essential for estimating the link flow from observed link speeds. Therefore, we can use the matrix relation to establish a direct connection between the observed speeds and the underlying traffic flow dynamics, leveraging the vast amount of GPS speed data and mobile devices. By integrating the assignment matrix $\mathbf{A}$ with the sparse traffic flow observations, we can formulate an optimization framework in AOR that accounts for both the spatial and temporal aspects of urban traffic flow.

\subsection{Analytical Formulation}

Building on the assignment matrix $\mathbf{A}$ provided in Section \ref{sec:assignment_mat}, a constrained linear optimization framework is developed to estimate both link flow and OD demand:

\begin{equation}\label{eq:l2_qua}
\begin{aligned}
& \min_{\mathbf{x},\mathbf{q}}||\mathbf{x}-\mathbf{A}\mathbf{q}||_{2}^{2}+w_{x}||\mathbf{x}||_{2}^{2}+w_{q}||\mathbf{q}||_{2}^{2} \\
& \text{s.t.} \quad \mathbf{x>0,}\mathbf{q}>0
\end{aligned}
\end{equation}

The objective function $\|\mathbf{x}-\mathbf{A}\mathbf{q}\|_{2}^{2}$ primarily seeks to minimize the least squares error, which quantifies the discrepancy between the estimated link flows $\mathbf{x}$, and the recovered flows derived from OD demands $\mathbf{q}$, using the assignment matrix $\mathbf{A}$. The regularization terms $\|\mathbf{x}\|_{2}^{2}$ and $\|\mathbf{q}\|_{2}^{2}$ control the magnitudes of the link flow vector $\mathbf{x}$ and the OD demand vector $\mathbf{q}$, respectively, helping to prevent overfitting. The weights $w_{x}$ and $w_{q}$ balance the influence of these regularization terms.

To refine the model, additional observed information such as partial traffic flow data from loop detectors or cameras, and long-term OD demand estimates from surveys or historical records, can be integrated. This is achieved by introducing adjustment terms, enhancing the model's realism:

\begin{equation}\label{eq:l2}
\begin{aligned}
\quad\quad\min_{\mathbf{x},\mathbf{q}} & ||\mathbf{x}-\mathbf{A}\mathbf{q}||_{2}^{2}+w_{x}||\mathbf{x}||_{2}^{2}+w_{q}||\mathbf{q}||_{2}^{2} +w_{sx}||M_{x}\mathbf{x}-\mathbf{x_{0}}||_{2}^{2}+w_{sq}||M_{q}\mathbf{q}-\mathbf{q_{0}}||_{2}^{2} \\
\text{s.t.} & \quad \mathbf{x>0,}\mathbf{q}>0
\end{aligned}
\end{equation}

Here, the matrix $M_{x}$ and $M_{q}$ are mapping matrices that indicate where traffic flow values are known. $\mathbf{x_{0}}$ and $\mathbf{q_{0}}$ represent the vectors of known flow and OD demand values, respectively. The observed link flow deviation penalty weight $w_{sx}$ and OD demands deviation penalty weight $w_{sq}$ are employed to balance the flow and demand recovery. The optimization of hyperparameters $w_{x}$, $w_{q}$, $w_{sx}$, $w_{sq}$ can be achieved through techniques like stochastic gradient descent. 

The differentiation of the objective function in Equation \ref{eq:l2} with respect to $\mathbf{x}$ and $\mathbf{q}$ transforms the quadratic objective function into a linear system. Setting the derivatives to zero yields key equations, which are then combined into a block matrix format representing the coefficient matrix and observed vectors. The linear system is given in Equation \ref{eq:l2_5}:

\begin{equation}\label{eq:l2_5}
\begin{bmatrix}\mathbf{x}\\
\mathbf{q}
\end{bmatrix}=\begin{bmatrix}(1+w_{x})I+w_{sx}M_{x}^{T}M_{x} & -\mathbf{A}\\
-\mathbf{A}^{T} & \mathbf{A}^{T}\mathbf{A}+w_{q}I+w_{sq}M_{q}^{T}M_{q}
\end{bmatrix}^{-1}\begin{bmatrix}w_{sx}M_{x}^{T}\mathbf{x_{0}}\\
w_{sq}M_{q}^{T}\mathbf{q_{0}}
\end{bmatrix}
\end{equation}

Let us denote the estimated vector by $\mathbf{E}$, the coefficient matrix by 
$\mathbf{M}$, and the observed vector by $\mathbf{V}$. Thus, we succinctly encapsulate the relationship expressed in Equation \ref{eq:l2_5} into a more streamlined form, shown in Equation \ref{eq:l2_6}:

\begin{equation}\label{eq:l2_6}
\mathbf{E}=\mathbf{M}^{-1}\mathbf{V}
\end{equation}

The inversion of the coefficient matrix $\mathbf{M}$ is essential but computationally intensive, especially for large, sparse matrices common in transportation network analyses. To address this, iterative algorithms such as the Conjugate Gradient (CG) method are used. The CG method is efficient for solving large-scale linear systems with symmetric and positive-definite matrices, minimizing the residual at each iteration and enhancing computational efficiency.

Although the current formulation does not incorporate non-negativity constraints, future developments in this methodology will introduce elements like Lagrange multipliers to address this, ensuring the feasibility of the derived solutions. This refinement will be detailed in subsequent subsections, improving the model's applicability to real-world traffic flow recovery tasks.

\subsection{SGD for Adjusting Hyperparameters} \label{sec:sgd}

In the optimization problem framework presented, several hyperparameters critically influence the optimization performance, as detailed in the accompanying table \ref{tab:hyper}.
\begin{table}[ht]
\caption{Hyperparameters}
\centering
\begin{tabular}{l c}
\hline \textbf{Hyperparameters} & \textbf{Math Notations} \\
\hline Link flow regularization weight & $w_{x}$ 	\\
Demand regularization weight & $w_{q}$ 	\\
Link flow deviation penalty weight & $w_{sx}$ 	\\
OD demands deviation penalty weight & $w_{sq}$ 	\\
 \hline
\end{tabular}
\label{tab:hyper}
\end{table}

These hyperparameters are instrumental in striking a balance between the model's fit to the data and its complexity. This balance is crucial for enhancing the accuracy and ensuring the generalizability of the model's estimations. To fine-tune these hyperparameters, the Stochastic Gradient Descent (SGD) algorithm is employed as an iterative method for iteratively adjusting hyperparameters to minimize the loss function, thereby refining the model's performance in a data-driven manner. 

The SGD procedure commences with the formulation of a loss function,$L$, meticulously designed to quantify the discrepancies between the estimated and observed link flows within the transportation network. This function specifically targets the subset of observed link flows, providing a reliable metric for assessing the model's estimation accuracy. The loss function is formulated as follows:

\begin{equation}\label{eq:l2_7}
L=\frac{1}{N_{x}\cdot H}||M_{x}\mathbf{x}-\mathbf{x_{0}}||_{2}^{2}
\end{equation}

Here, $N_{x}$ represents the number of observed data points in the link flow dataset, while H denotes the number of time intervals within the planning horizon. The mapping matrix $M_{x}$ is utilized to select corresponding elements from the estimated link flow vector $\mathbf{x}$ for direct comparison with the observed link flow dataset $x_{0}$. To optimize each hyperparameter $w_{i}$, we calculate the gradient of the loss function $L$ with respect to these parameters, using the chain rule: 

\begin{equation}\label{eq:l2_8}
\frac{\partial L}{\partial w_{i}}=\frac{\partial L}{\partial\mathbf{x}}\frac{\partial\mathbf{x}}{\partial w_{i}}
\end{equation}

The chain rule here allows us to decompose the derivative of the loss function's gradient into more manageable parts, such as the linear system defined in above section. The derivative of the loss function $L$ with respect to the estimated link flow vector $\mathbf{x}$ is computed by differentiating L with respect to $\mathbf{x}$:

\begin{equation}\label{eq:l2_9}
\frac{\partial L}{\partial\mathbf{x}}=2\frac{1}{N_{x}\cdot H}M_{x}^{T}(M_{x}\mathbf{x}-\mathbf{x_{0}})
\end{equation}

To compute the $\frac{\partial\mathbf{x}}{\partial w_{i}}$ for each hyperparameter, we refer to the structure of the linear analytical formulation established in Equation \ref{eq:l2_5}. Therefore, the derivatives of link flow vector $\mathbf{x}$ with respect to each hyperparameters $w_{x}$, $w_{q}$,$w_{sx}$ and $w_{sq}$ are calculated as follows:

\begin{equation}\label{eq:l2_11}
\frac{\partial\mathbf{x}}{\partial w_{x}}=-\mathbf{M}^{-1}\begin{bmatrix}I & 0\\
0 & 0
\end{bmatrix}\mathbf{M}^{-1}\mathbf{V}
\end{equation}

\begin{equation}\label{eq:l2_12}
\frac{\partial\mathbf{x}}{\partial w_{q}}=-\mathbf{M}^{-1}\begin{bmatrix}0 & 0\\
0 & I
\end{bmatrix}\mathbf{M}^{-1}\mathbf{V}
\end{equation}

\begin{equation}\label{eq:l2_13}
\frac{\partial\mathbf{x}}{\partial w_{sx}}=-\mathbf{M}^{-1}\begin{bmatrix}M_{x}^{T}M_{x} & 0\\
0 & 0
\end{bmatrix}\mathbf{M}^{-1}\mathbf{V}+\mathbf{M}^{-1}\begin{bmatrix}M_{x}^{T}X_{0}\\
0
\end{bmatrix}
\end{equation}

\begin{equation}\label{eq:l2_14}
\frac{\partial\mathbf{x}}{\partial w_{sq}}=-\mathbf{M}^{-1}\begin{bmatrix}0 & 0\\
0 & M_{q}^{T}M_{q}
\end{bmatrix}\mathbf{M}^{-1}\mathbf{V}+\mathbf{M}^{-1}\begin{bmatrix}0\\
M_{q}^{T}q_{0}
\end{bmatrix}
\end{equation}
After obtaining these derivatives, we can update each old hyperparameter $w_{i}^{old}$ using the gradient descent update rule with learning rate $\alpha$:

\begin{equation}\label{eq:l2_15}
w_{i}^{new}=w_{i}^{old}-\alpha\frac{\partial L}{\partial w_{i}}
\end{equation}

For effective SGD implementation, a certain dataset of traffic observations taken at different times within the same transportation network is essential. The iterative process proceeds until all the collected samples have been utilized, ensuring comprehensive hyperparameter optimization. The SGD procedure, as delineated in Algorithm \ref{alg:sgd}, methodically refines the hyperparameters.

\begin{algorithm}
\caption{Stochastic Gradient Descent (SGD) for adjusting hyperparameters}\label{alg:sgd}
\begin{algorithmic}[1]
\STATE \textbf{Input:} Traffic data samples $\{\mathbf{X_{1},...,X_{n}}\}$, loss function $L$, learning rate $\alpha$, number of epochs $N$
\STATE \textbf{Output:} Updated hyperparameters $\{w_{x},w_{q},w_{sx},w_{sq}\}$
\STATE Initialize hyperparameters $\{w_{x}^{0},w_{q}^{0},w_{sx}^{0},w_{sq}^{0}\}$ randomly
    \FOR{$j\leftarrow1$ to $N$} 
    \STATE Conduct analytical method on current randomly selected traffic data sample $\mathbf{X}_{j}$ to obtain coefficient matrix $\mathbf{M_{j}}$ and observed vector $\mathbf{V_{j}}$
    \STATE Compute the derivative of the loss function $L_{j}$ with respect to the estimated link flow vector $\mathbf{x}$
    \STATE Compute the derivative of the link flow vector $\mathbf{x_{j}}$ with respect to current each hyperparameters $w_{x}^{j}, w_{q}^{j},w_{sx}^{j}$ and $w_{sq}^{j}$
    \STATE Utilize the chain rule to compute the loss function $L_{j}$ with respect to each hyperparameters $w_{x}^{j}, w_{q}^{j},w_{sx}^{j} $ and $ w_{sq}^{j}$
    \STATE Update hyperparameters $w_{i}^{j+1}=w_{i}^{j}-\alpha\frac{\partial L}{\partial w_{i}^{j}}$
    \ENDFOR
\end{algorithmic}
\end{algorithm}

\subsection{Lagrange Relaxation}\label{sec:lr}

To address the non-negativity constraint $\mathbf{x > 0, q > 0}$, a common approach in previous work resorts to straightforwardly projecting negative values obtained post-solution to zero \citep{ma2018estimating}. This practice, while pragmatic, risks masking inherent inaccuracies in the modeling process or discrepancies within the data itself. To circumvent these issues, our formulation introduces the application of a Lagrange Relaxation method. This approach effectively incorporates the non-negativity constraints into the objective function through the multiplication of constraints by their respective Lagrange multipliers. Consequently, the Lagrangian, denoted as $Z(\mathbf{x, q}, \lambda_{x}, \lambda_{q})$, is constructed in Equation \ref{eq:l2_lag}, seamlessly integrating these constraints within the optimization framework and preserving the integrity of the modeling process.

\begin{equation}\label{eq:l2_lag}
\begin{aligned}
\min_{\mathbf{x}, \mathbf{q}} & \ ||x - Aq||_{2}^{2} + w_{x}||x||_{2}^{2} + w_{q}||q||_{2}^{2} + w_{sx}||M_{x}x - x_{0}||_{2}^{2} + w_{sq}||M_{q}q - q_{0}||_{2}^{2} \\
& \quad - \sum \lambda_{x,i} x_{i} - \sum \lambda_{q,j} q_{j}
\end{aligned}
\end{equation}

Here, $\lambda_{x,i}$ and $\lambda_{q,j}$ are the components of the Lagrange multiplier vectors $\mathbf{\lambda_{x}}$ and $\mathbf{\lambda_{q}}$, respectively, and $x_{i}, q_{j}$ are components of vectors $\mathbf{x}$ and $\mathbf{q}$. Therefore, the optimization problem becomes finding the values of $\mathbf{x},\mathbf{q},\mathbf{\lambda_{x}}$ and $\mathbf{\lambda}_{q}$ that minimize the Lagrangian. This method allows for a more nuanced solution that respects the non-negativity constraints while providing a flexible framework for handling these constraints compared to simple projection methods.

To address the challenges of enforcing non-negativity constraints in traffic flow estimation for large-scale transportation networks, our methodology employs a refined Lagrange Relaxation (LR) approach. This method simplifies the optimization problem by transforming these constraints into penalty terms within the objective function. Central to this approach is the integration of Lagrange multiplier vectors $\lambda_{x}$ and $\lambda_{q}$, as detailed in Section 3.2 of our methodology. These multipliers are iteratively adjusted to penalize and progressively eliminate any violations of the non-negativity constraints on the estimated link flows and origin-destination demands.

The designed Lagrange relaxation approach involves an iterative process to adjust the Lagrange multipliers vectors at each iteration, the essence of this approach lies in evaluating the gap between two solution states: the current relaxed solution 
$Z_{R}^{t}(\mathbf{x^{t},q^{t}},\lambda_{x}^{t},\lambda_{q}^{t})$, and the current feasible solution $Z_{F}^{t}(\mathbf{x_{f}^{t}}$,$\mathbf{q_{f}^{t}})$. The relaxed solution $Z_{R}^{t}(\mathbf{x^{t},q^{t}},\lambda_{x}^{t},\lambda_{q}^{t})$ is computed without incorporating the non-negativity constraints on $\mathbf{x}$ and $\mathbf{q}$ but with the current values of $\lambda_{x}^{t}$ and $\lambda_{q}^{t}$. It is defined by the following function:

\begin{equation}\label{eq:l2_16}
\begin{aligned}
Z_{R}^{t}(\mathbf{x^{t},q^{t}},\lambda_{x}^{t},\lambda_{q}^{t}) &= ||\mathbf{x}^{t}-\mathbf{A}\mathbf{q}^{t}||_{2}^{2} + w_{x}||\mathbf{x}^{t}||_{2}^{2} + w_{q}||\mathbf{q}^{t}||_{2}^{2} \\
&\quad + w_{sx}||M_{x}\mathbf{x}^{t}-\mathbf{x_{0}}||_{2}^{2} + w_{sq}||M_{q}\mathbf{q}^{t}-\mathbf{q_{0}}||_{2}^{2} \\
&\quad -\lambda_{x}^{t}\mathbf{x}^{t} -\lambda_{q}^{t}\mathbf{q}^{t}
\end{aligned}
\end{equation}

The feasible solution $Z_{F}^{t}(\mathbf{x_{f}^{t}},\mathbf{q_{f}^{t}})$, on the other hand, respects the non-negativity constraints by applying a maximum operation to ensure that both $\mathbf{x}$ and $\mathbf{q}$ are non-negative. It is defined as:
\begin{equation}\label{eq:l2_17}
\mathbf{x_{f}^{t}}=max(0,\mathbf{x^{t}}),\mathbf{q_{f}^{t}}=max(0,\mathbf{q^{t}})
\end{equation}
\begin{equation}\label{eq:l2_18}
Z_{F}^{t}(\mathbf{x_{f}^{t},q_{f}^{t}})=||\mathbf{x_{f}}^{t}-\mathbf{A}\mathbf{q_{f}}^{t}||_{2}^{2}+w_{x}||\mathbf{x_{f}}^{t}||_{2}^{2}+w_{q}||\mathbf{q_{f}}^{t}||_{2}^{2}+w_{sx}||M_{x}\mathbf{x_{f}}^{t}-\mathbf{x_{0}}||_{2}^{2}+w_{sq}||M_{q}\mathbf{q_{f}}^{t}-\mathbf{q_{0}}||_{2}^{2}
\end{equation}
To compute these solutions, the matrix equations to get estimated values are adapted as follows:
\begin{equation}\label{eq:l2_19}
\begin{bmatrix}\mathbf{x^{t}}\\
\mathbf{q^{t}}
\end{bmatrix}=\begin{bmatrix}(1+w_{x})I+w_{sx}M_{x}^{T}M_{x} & -\mathbf{A}\\
-\mathbf{A}^{T} & \mathbf{A}^{T}\mathbf{A}+w_{q}I+w_{sq}M_{q}^{T}M_{q}
\end{bmatrix}\begin{bmatrix}w_{sx}M_{x}^{T}\mathbf{x_{0}}+\lambda_{x}^{t}\\
w_{sq}M_{q}^{T}\mathbf{q_{0}}+\lambda_{q}^{t}
\end{bmatrix}
\end{equation}

This modification introduces Lagrange multiplier vectors into the right-hand constant vectors of the system, allowing to evaluate both the relaxed and feasible solutions. The gap of relaxed solution and feasible solution informs the update of the Lagrange multipliers, propelling the solution towards feasibility. The iterative update process involves a carefully calculated step size $s^{t}$, which is dynamically adjusted based on the magnitude of the gap between the relaxed and feasible solutions. The step size formula is given by:
\begin{equation}\label{eq:l2_20}
s^{t}=\frac{\mu^{t}(Z_{R}^{t}(\mathbf{x^{t},q^{t}},\lambda_{x}^{t},\lambda_{q}^{t})-Z_{F}^{t}(\mathbf{x_{f}^{t}},\mathbf{q_{f}^{t}}))}{(\sum_{i=1}^{n}\mathbf{x_{i}}^{t}+\sum_{j=1}^{n}\mathbf{q_{j}}^{t})^{2}}
\end{equation}
Here, $\mu^{t}$ is a control parameter that is adjusted based on the convergence of the gap. Specifically, if the gap does not show significant improvement over a set number of iterations, decrease $\mu^{t}$ by a factor of 2. The terms $x_{i}^{t}$ and $q_{j}^{t}$ are components of the current estimated vectors $\mathbf{x}^{t}$ and $\mathbf{q^{t}}$, respectively. The denominator $(\sum_{i=1}^{n}x_{i}^{t}+\sum_{j=1}^{n}q_{j}^{t})^{2}$ normalizes the step size to the scale of the estimated link flows and OD demands, ensuring that the updates are proportionate to the size of the variables involved. With each iteration, the Lagrange multipliers are updated according to the computed step size as follows:
\begin{equation}\label{eq:l2_21}
\lambda_{x}^{t+1}=max\{0,\lambda_{x}^{t}+s^{t}\mathbf{x}^{t}\}
\end{equation}
\begin{equation}\label{eq:l2_22}
\lambda_{q}^{t+1}=max\{0,\lambda_{q}^{t}+s^{t}\mathbf{q}^{t}\}
\end{equation}
This process reduces the penalty for constraint violations if slack exists in the relaxed constraint. Conversely, it intensifies the penalty to diminish such violations. This iterative method proceeds until the solution adequately converges within a feasible domain, ensuring compliance with the non-negativity constraints while optimizing the traffic flow estimation for large-scale transportation networks. The procedure of the LR approach is presented in Algorithm \ref{alg:lr}:

\begin{algorithm}
\caption{Lagrange relaxation approach}\label{alg:lr}
\begin{algorithmic}[1]
\STATE \textbf{Input:} Matrices and vectors $\{\mathbf{A},M_{x},M_{q},\mathbf{x_{0}},\mathbf{q_{0}}\}$,hyperparameters $\{w_{x},w_{q},w_{sx},w_{sq}\}$, control parameter $\mu^{t}$, number of iterations $N$, gaps threshold $G_{T}$
\STATE \textbf{Output:} Lagrange multiplier vectors $\{\lambda_{x}^{N},\lambda_{q}^{N}\}$,estimated feasible solutions $\mathbf{x_{f}^{N}}$
\STATE Initialize Lagrange multiplier vectors $\lambda_{x}^{0}$ and $\lambda_{q}^{0}$ as zero vectors
\FOR{$t\leftarrow1$ to $N$}
    \STATE Compute the estimated vectors $\mathbf{x^{t}}$ and $\mathbf{q^{t}}$, and their feasible counterparts $\mathbf{x_{f}^{t}}$ and $\mathbf{q_{f}^{t}}$, using the analytical method
    \STATE Compute the current relaxed solution $Z_{R}^{t}(\mathbf{x^{t},q^{t}},\lambda_{x}^{t},\lambda_{q}^{t})$ and current feasible solution $Z_{F}^{t}(\mathbf{x_{f}^{t}},\mathbf{q_{f}^{t}})$
    \STATE Compute the current gaps $G^{t}=Z_{R}^{t}(\mathbf{x^{t},q^{t}},\lambda_{x}^{t},\lambda_{q}^{t})-Z_{F}^{t}(\mathbf{x_{f}^{t}},\mathbf{q_{f}^{t}})$
    \IF{$|G^{t}-G^{t-1}|<G_{T}$}
        \STATE $\mu^{t}\leftarrow\mu^{t}/2$
    \ELSE
        \STATE $\mu^{t}\leftarrow\mu^{t}$
    \ENDIF
    \STATE Update step size $s^{t}\leftarrow\frac{\mu^{t}\cdot G^{t}}{(\sum_{i=1}^{n}\mathbf{x_{i}}^{t}+\sum_{j=1}^{n}\mathbf{q_{j}}^{t})^{2}}$
    \STATE Update Lagrange multiplier vectors $\lambda_{x}^{t+1}\leftarrow max\{0,\lambda_{x}^{t}+s^{t}\mathbf{x}^{t}\}$ and $\lambda_{q}^{t+1}\leftarrow max\{0,\lambda_{q}^{t}+s^{t}\mathbf{q}^{t}\}$
\ENDFOR
\end{algorithmic}
\end{algorithm}

\section{Numerical Experiments}
This section evaluates the proposed methodology using hypothetical scenarios set within a large-scale road network in Shenzhen's Futian District. These case studies are carefully designed to assess the methodology's performance across different road types within the network. Given the challenges of obtaining comprehensive city-level speed data in such a large network, simulated traffic data generated by the SUMO platform is used as a substitute for actual traffic conditions, allowing for a thorough analysis of the methodology's capabilities. Additionally, the methodology is tested on a smaller road network using real-world traffic data to further validate its effectiveness. Together, these studies demonstrate the practicality of the AOR approach under varying traffic dynamics and network complexities. The following sections provide a detailed examination of the methodology's application in these scenarios. The experiments below are conducted on a desktop with AMD Ryzen 9 3950X CPU@ 3.49 GHz \texttimes{} 16, 32 GB RAM, 1 TB SSD.

\subsection{Dataset Description}
This section describes the characteristics of the transportation network and the traffic data employed in the hypothetical case study.

The case study focuses on the Futian District of Shenzhen, Guangdong Province, China. Spanning an area of ${78.8 \mathrm{~km}^2}$, detailed characteristics of the road network are presented in Table \ref{tab:network_setting_hypo}. 

\begin{table}[ht]
\caption{Network characteristics}
\centering
\begin{tabular}{l l}
\hline \textbf{Parameters} & \textbf{values} \\
\hline Road Network Areas & ${78.8 \mathrm{~km}^2}$ \\
 Number of Road Segments & ${9,409}$ \\
 Number of Junctions & ${6,754}$ \\
 Highways & 41 \\
 Expressways & ${185}$ \\
 Arterial Roads & ${1,399}$ \\
 Secondary Roads & ${1,249}$ \\
 Branch Roads & ${4,532}$ \\
 Frontage Roads & ${1,200}$ \\
 Interchange Ramp & 803 \\
\hline
\end{tabular}
\label{tab:network_setting_hypo}
\end{table}

\begin{figure*}[ht]
    \centering
    \includegraphics[width=0.8\textwidth]{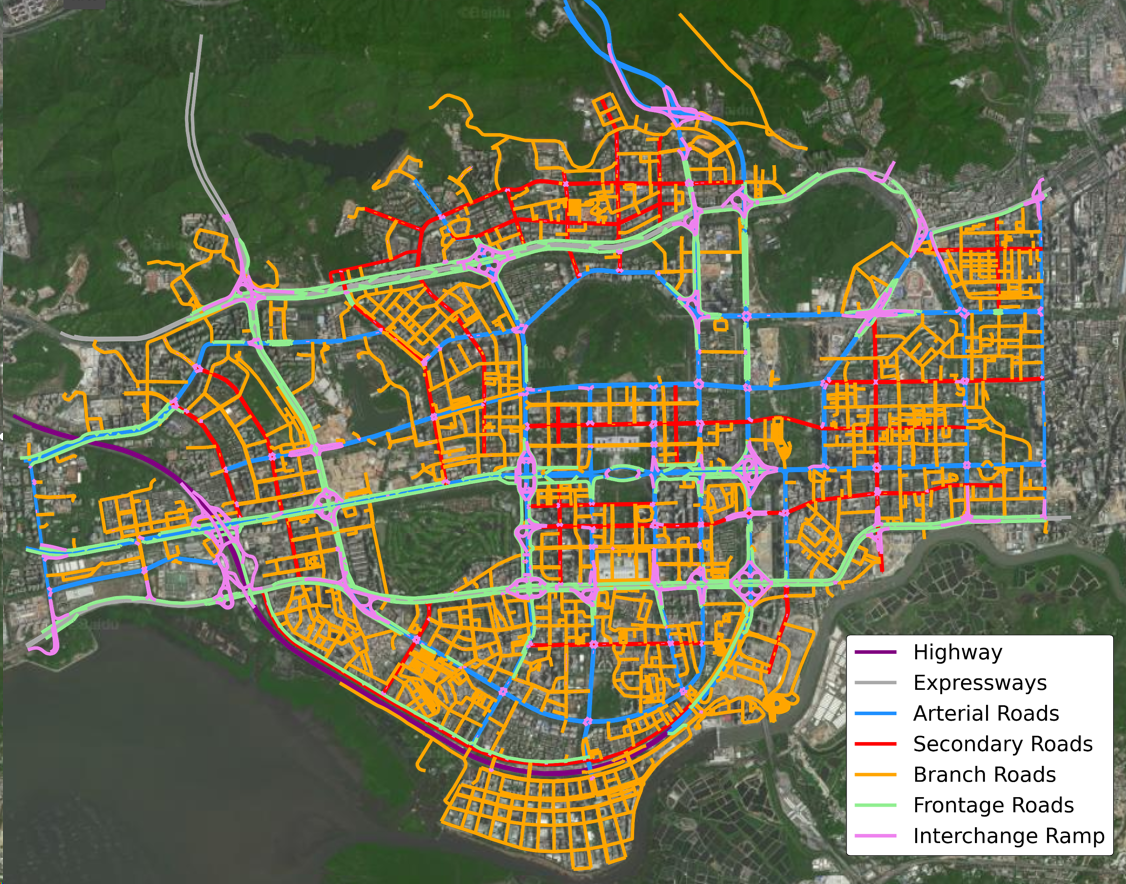}
    \caption{GIS representation of the road network in Futian District}
    \label{fig:GIS_view_hypo}
\end{figure*}

This expansive network mirrors the complexity of urban transportation infrastructure and offers an extensive overview of various traffic conditions and road features, ensuring a thorough examination of transportation dynamics. The GIS view of the road network and the specific road type distribution in the Futian District are presented in Figure \ref{fig:GIS_view_hypo}. 

\begin{figure*}[ht]
\centering
\begin{minipage}{0.45\textwidth}
    \centering
    \includegraphics[width=\linewidth]{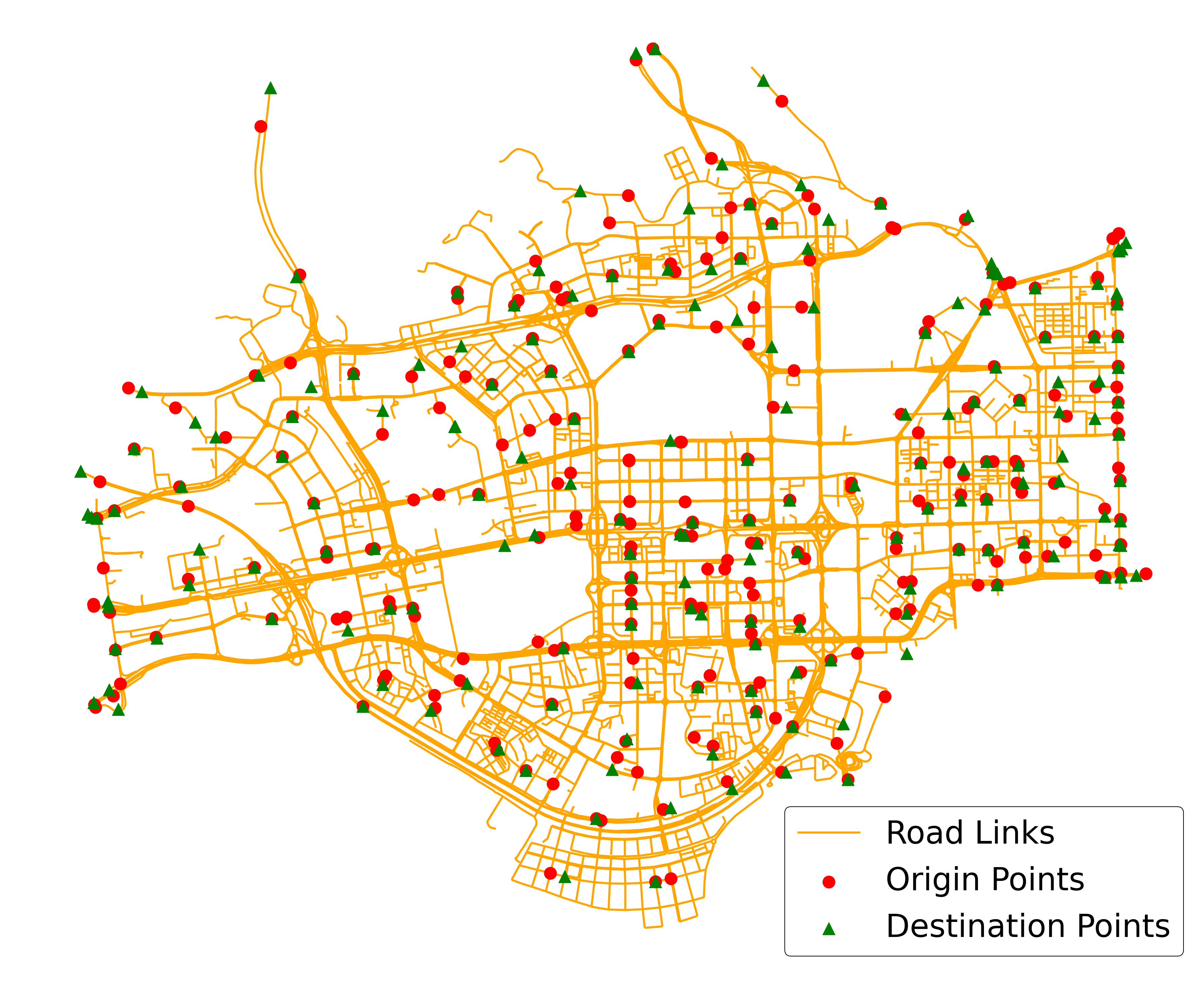}
    \caption{Spatial distribution of OD points}
    \label{fig:road_OD}
\end{minipage}%
\hfill
\begin{minipage}{0.45\textwidth}
    \centering
    \includegraphics[width=\linewidth]{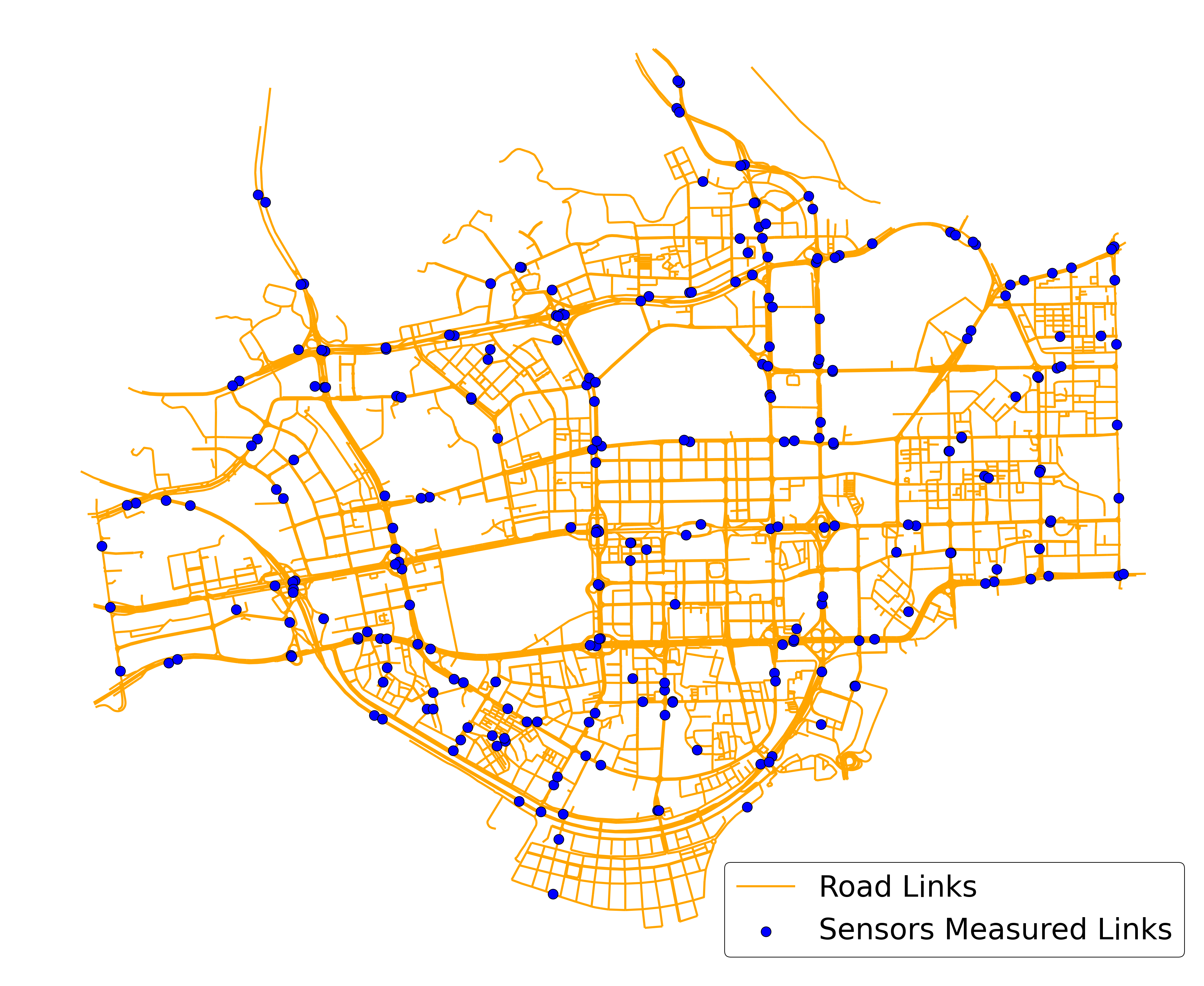}
    \caption{Traffic surveillance camera locations in Futian District}
    \label{fig:camera_dist}
\end{minipage}
\end{figure*}

Before initiating the simulation, it is imperative to specify the OD points and the associated demand data. Given the challenges of collecting detailed OD data for extensive networks, the Shenzhen Department of Transportation collected verifiable OD demand information through floating car trajectories, travel surveys, and a population model based on land use across the Futian District. This data was collected during a specific one-hour timeframe on August 1st, 2021, focusing on three critical time intervals: the morning peak hour (8:00 to 9:00 AM), midday (12:00 to 1:00 PM), and the evening peak hour (6:00 to 7:00 PM). The OD demand information comprises 2,382 OD pairs based on the actual travel patterns of Futian residents. The spatial distribution and arrangements of origin and destination points are shown in Figure \ref{fig:road_OD}. This OD information is employed in the model to determine eligible paths for each OD pair, enhancing the analysis of traffic dynamics.

Identifying optimal locations for sensor deployment to achieve effective traffic flow recovery is a complex challenge. At the current stage of our research, we utilize the existing distribution of surveillance cameras within the Futian District, provided by the Shenzhen Department of Transportation. This approach involves generating an observed link flow vector that reflects the actual sensor locations, as depicted in Figure \ref{fig:camera_dist}. In total, sensors monitor 218 links, covering only 2.32\% of the entire network. The traffic surveillance cameras are primarily concentrated on highways, expressways, and arterial roads, reflecting their higher traffic demand. However, camera coverage on branch roads remains low, highlighting the prioritization of monitoring high-traffic areas and the challenges in achieving comprehensive surveillance across all road types.

Using the observed OD data, a simulation environment was constructed within SUMO to closely mimic real-world traffic conditions. SUMO is a versatile, open-source traffic simulation software that models and analyzes various transportation scenarios. In this study, SUMO was used to generate ground-truth speed and traffic flow data across various road segments in the Futian District's road network, simulating traffic dynamics over a one-hour duration aligned with the observed OD demand periods. Traffic flow and average speed data were gathered using simulated loop detectors placed throughout the road network, collecting data in 5-minute intervals for each lane. Data from individual lanes were then aggregated or averaged at the road link level. The dataset for Futian's road network includes detailed information on average speed and traffic flow for each road link, compiled in 5-minute segments over one hour, resulting in 12-time intervals per link. This comprehensive dataset is crucial for evaluating the effectiveness of the proposed AOR method in managing traffic within a large-scale urban transportation environment, providing a rigorous assessment of its performance under controlled conditions.

\subsection{Evaluation Metrics}\label{sec:eval_metrics}

As aforementioned, there are four penalty weights in the linear
optimization equation, which can noticeably affect the proposed model performance. An evaluation metric is necessary to assess the recovery of road network flow performance and penalty weight value selection.
Because the traffic flow varies greatly on different road links, relying solely on Mean Absolute Error (MAE) and Relative Mean Error (RME) may not provide accurate measurements performance for modeling every link in the road network \cite{meng2017city,zhang2020network}. Numerous studies have adopted the Mean Absolute Percentage Error (MAPE) as a metric for evaluating the performance of traffic flow estimations. However, the MAPE formulation encounters limitations, particularly when addressing conditions where traffic flow may be zero, rendering it less effective for comprehensive error analysis across all traffic conditions. In response to this limitation and with an emphasis on accurately assessing high-demand traffic roads, this analysis introduces a Weighted Relative Mean Error (WRME) function. This refined metric is designed to compute the error by incorporating weights that reflect the varying significance of different roads, thereby offering a more nuanced and effective approach to error calculation in overall traffic flow recovery. The WRME function is defined through equations \ref{eq:Weight} and \ref{eq:WRME} to quantify the discrepancy between observed and recovered traffic flows. The weight $W_{l}$ for each link $l$ is calculated as the ratio of the total observed flow on link $l$ across all time periods $t$ to the aggregate flow observed across the entire network, as expressed in Equation \ref{eq:Weight}. This weighting approach ensures that the significance of each link's flow is proportionally represented in the overall error calculation.

\begin{equation}
W_{l}=\frac{\sum_{t=t_{0}}^{T}{f_{l}^{t}}}{\sum_{l=l_{0}}^{L}\sum_{t=t_{0}}^{T}{f_{l}^{t}}}
\label{eq:Weight}
\end{equation}

Subsequently, the WRME, delineated in Equation \ref{eq:WRME}, aggregates these weights alongside the relative mean error of each link to compute a holistic measure of estimation accuracy across the network. This metric specifically addresses the criticality of accurately estimating flows on high-demand links, where discrepancies can have pronounced impacts on traffic management and analysis.

\begin{equation}
WRME=\sum_{l=l_{0}}^{L}(W_{l}*\frac{\sum_{t=t_{0}}^{T}|f_{l}^{t}-\hat{f_{l}^{t}}|}{\sum_{t=t_{0}}^{T}f_{l}^{t}})
\label{eq:WRME}
\end{equation}

Where $\hat{f_{l}^{t}}$ represents the estimated traffic flow passing through link $l$ during time interval $t$, while the ground-truth observed flow value of link $l$ during time period $t$ is represented by$f_{l}^{t}$. $T$ denotes the total number of observation time intervals, while $L$ refers to the total number of traffic assigned link sets. The WRME thus aims to refine the accuracy of flow recovery, particularly for links with high traffic volume, by minimizing the weighted error across the network. 

Furthermore, to assess the recovery performance on an individual link basis, it is imperative to implement the Mean Absolute Error (MAE) and Relative Mean Error (RME) metrics for each link. These measures are essential for delving into the effectiveness of recovery on specific links within the network. The respective equations for RME and MAE, as delineated in Equations \ref{eq:RME} and \ref{eq:MAE}, facilitate a focused evaluation of each link's recovery accuracy.

\begin{equation}
RME_l=\frac{\sum_{t=t_{0}}^{T}|f_{l}^{t}-\hat{f_{l}^{t}}|}{\sum_{t=t_{0}}^{T}f_{l}^{t}}
\label{eq:RME}
\end{equation}

\begin{equation}
MAE_l=\frac{1}{T}\sum_{t=t_{0}}^{T}|f_{l}^{t}-\hat{f_{l}^{t}}|
\label{eq:MAE}
\end{equation}

In these formulations, $RME_l$ computes the proportion of the absolute error in traffic flow estimation relative to the actual flow for link $l$ across all observed time intervals $T$, offering a normalized measure of estimation error. Conversely, $MAE_l$ quantifies the average magnitude of error per time interval, providing an absolute measure of estimation accuracy for the link. Both metrics, by quantifying discrepancies between the estimated traffic flow $\hat{f_{l}^{t}}$ and the ground-truth observed flow $f_{l}^{t}$, furnish critical insights into the specific recovery effects and efficacy of the estimation process for individual traffic links.

\subsection{Parameters Settings}
For smaller networks, it is feasible to enumerate all possible paths. However, in larger city-level traffic networks, the K shortest paths for each OD pair are enumerated to balance computational efficiency with comprehensive network coverage. A K-shortest paths algorithm based on travel times generates routes for given OD pairs, with $k$ set to 5.

\begin{figure}[ht]
\centering
\includegraphics[width=0.55\textwidth]{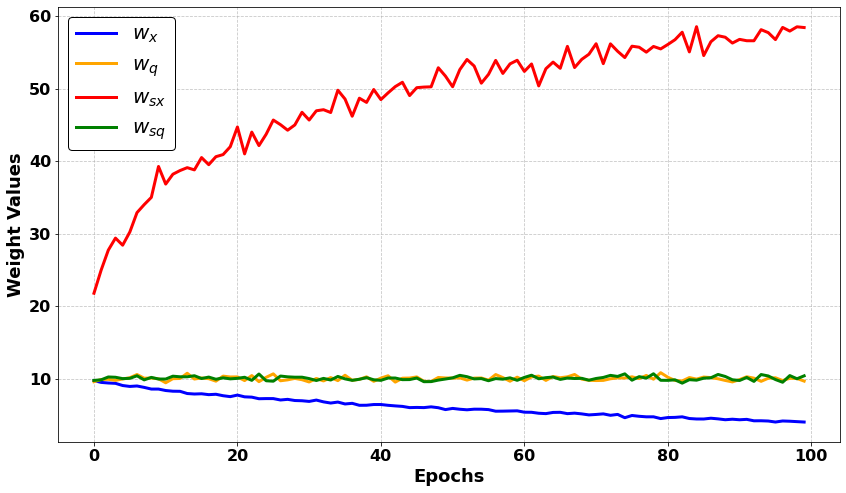}
\caption{Evolution of hyperparameters in the Stochastic Gradient Descent (SGD) process}
\label{fig:sgd}
\end{figure}

An SGD process, described in Section \ref{sec:sgd}, is adopted for iterative fine-tuning. This process involves setting the number of epochs to 100 and the learning rate to 0.01. Data from 24 hourly samples on August 1st, 2021, are randomly selected for each iteration to progressively refine the model parameters. The evaluation of four key hyperparameters through the SGD process is presented in Figure \ref{fig:sgd}, which reveals that the hyperparameter $w_{sx}$ exhibits a growth trend through successive epochs, indicating its increasing significance in the optimization process. Conversely, the hyperparameter $w_{x}$ displays a decreasing trajectory, suggesting that the regularization term $||\mathbf{x}||_2$ plays a diminishing role as the model evolves. This leads to adopting parameter values at the 100th epoch as near-optimal hyperparameters for subsequent evaluations of the model's recovery performance. Figure \ref{fig:sgd} reveals that the hyperparameter $w_{sx}$ exhibits a growth trend through successive epochs, indicating an increasing significance in the optimization process. Conversely, the hyperparameter $w_{x}$ displays a decreasing trajectory, suggesting that the regularization term $||\mathbf{x}||_2$ plays a diminishing role as the model evolves. This observation leads to the adoption of parameter values at the 100th epoch as near-optimal hyperparameters, providing a refined basis for subsequent evaluations of the model's recovery performance.

\begin{figure}[ht]
    \centering
    \begin{minipage}{0.3\textwidth}
        \centering
        \includegraphics[width=\linewidth]{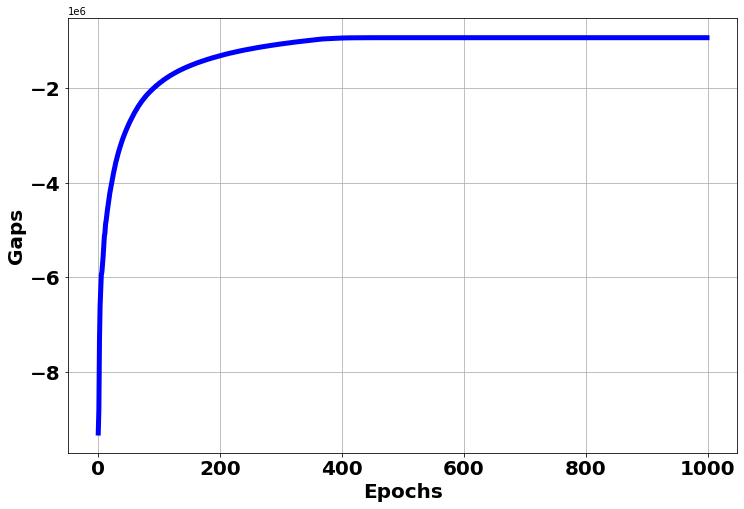}
        \caption*{(a) Gaps}
        \label{fig:lr1}
    \end{minipage}%
    \hfill
    \begin{minipage}{0.3\textwidth}
        \centering
        \includegraphics[width=\linewidth]{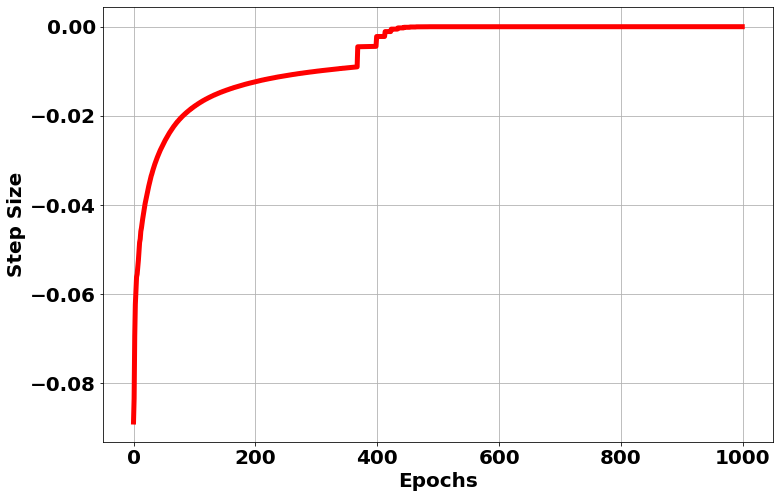}
        \caption*{(b) Step size}
        \label{fig:lr2}
    \end{minipage}%
    \hfill
    \begin{minipage}{0.3\textwidth}
        \centering
        \includegraphics[width=\linewidth]{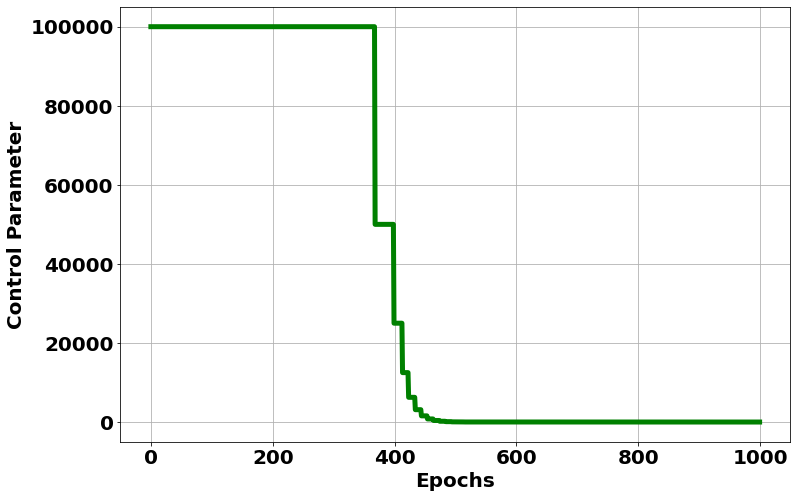}
        \caption*{(c) Control parameters}
        \label{fig:lr3}
    \end{minipage}
    \caption{LR process in the framework}
    \label{fig:combined_sgd}
\end{figure}

In addition, the LR method is executed with an initial control parameter, $\mu^t$ set at $1\times 10^5$ and a default gap threshold $G_T$ of $500$. The LR process is extended over 1000 epochs to monitor the model's convergence. The progression of the LR method, applied to PM peak data, is illustrated in Figure \ref{fig:combined_sgd}. 

Specifically, Figure \ref{fig:combined_sgd}(a) shows the convergence of gap values toward zero as the number of epochs increases, indicating the optimization's progress toward feasibility. Figure \ref{fig:combined_sgd}(b) demonstrates the adaptive nature of the step size, which increases over the epochs in response to the optimization landscape. Finally, Figure \ref{fig:combined_sgd}(c) tracks the evolution of the control parameters throughout the epochs, highlighting the dynamic adjustment process in the search for optimal solutions. The convergence around the 500-epoch mark, as depicted in these figures, underscores the efficacy of the LR method in iteratively refining the multiplier vectors $\lambda_x$ and $\lambda_q$. Achieving near-optimal values for these vectors at the convergence point helps align the recovered traffic flows with actual observable data while maintaining the model's physical plausibility.

\subsection{Experiment Analysis}
To evaluate the overall network performance, the WRME was computed to assess the flow recovery during the AM peak, midday, and PM peak periods. The results of this assessment are encapsulated in Table \ref{tab:l2_performance_metrics}.

\begin{table}[ht]
\caption{Recovery performance metrics for different periods}
\centering
\begin{tabular}{l c c c}
\hline 
\textbf{Metrics} & \textbf{AM Peak} & \textbf{Midday} & \textbf{PM Peak} \\
\hline 
Observation time & 08:00 - 09:00 & 12:00 - 13:00 & 18:00 - 19:00 \\
Average solving time & 1min 37s & 1min 28s & 1min 42s \\
Average WRME & \textbf{0.1984} & \textbf{0.1837} & \textbf{0.1606} \\
\hline
\end{tabular}
\label{tab:l2_performance_metrics}
\end{table}

The table indicates that the model was able to process the data within an average solving time ranging from 1 minute 28 seconds to 1 minute 42 seconds for the different time slots. Notably, the model achieved the lowest average WRME of 0.1606 during the PM peak period. A further performance evaluation by road type underscores the effectiveness of the techniques implemented in the AOR framework, as detailed in Table \ref{tab:l2_road_WRME}. Highways, expressways, and arterial roads, which have higher traffic demand and more camera deployments, show lower error rates. Conversely, road types with lower demand, such as branch roads, frontage roads, and interchange ramps, exhibit higher error rates but still maintain good fitness. This indicates that the AOR framework is particularly effective for high-demand roads while also performing reasonably well across different road types.

\begin{figure*}[ht]
\centering
\includegraphics[width=0.65\textwidth]{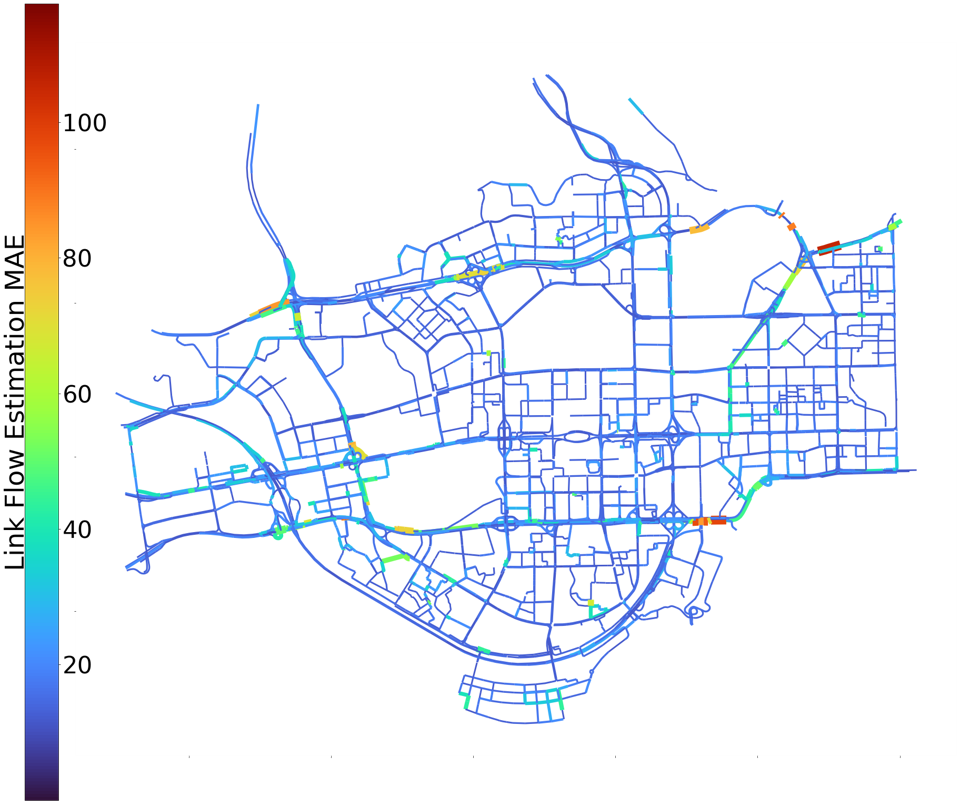}
\caption{Network-wide visualization of Mean Absolute Error (MAE) distribution}
\label{fig:l2_mae_perform}
\end{figure*}

\begin{table}[ht]
\caption{Recovery WRME performance by road type}
\centering
\begin{tabular}{l l}
\hline \textbf{Road types} & \textbf{WRME} \\
\hline 
 Highways & ${0.0300}$ \\
 Expressways & ${0.0537}$ \\
 Arterial Roads & ${0.1264}$ \\
Secondary Roads & ${0.2929}$ \\
 Branch Roads & ${0.4002}$ \\
 Frontage Roads & ${0.4627}$ \\
 Interchange Ramps & ${0.2463}$ \\
\hline
\end{tabular}
\label{tab:l2_road_WRME}
\end{table}

For a detailed assessment of the recovery performance, the MAE was calculated for each link across the network. Figure \ref{fig:l2_mae_perform} shows the MAE distribution, indicating strong overall recovery performance with minimal discrepancies, except in a few cases due to geographical peculiarities. This analysis highlights the accuracy achieved through the AOR framework, demonstrating its effectiveness in minimizing estimation errors across a diverse urban network on a macroscopic level.

\begin{figure}[ht]
    \centering
    \begin{minipage}{0.45\textwidth}
        \centering
        \includegraphics[width=\textwidth]{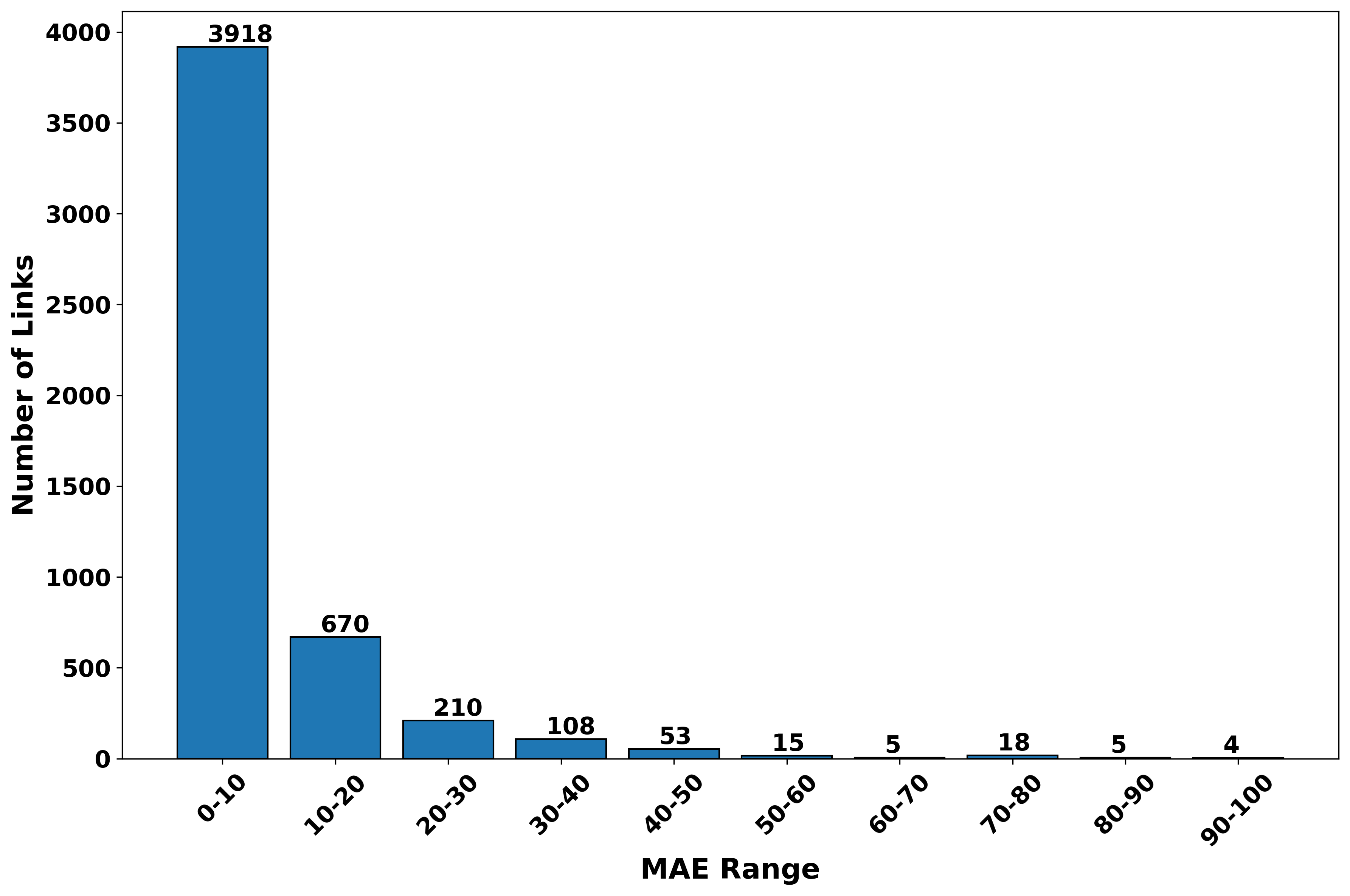}
        \caption{Histogram plot of the distribution of links by MAE ranges}
        \label{fig:mae_dist}
    \end{minipage}\hfill
    \begin{minipage}{0.45\textwidth}
        \centering
        \includegraphics[width=\textwidth]{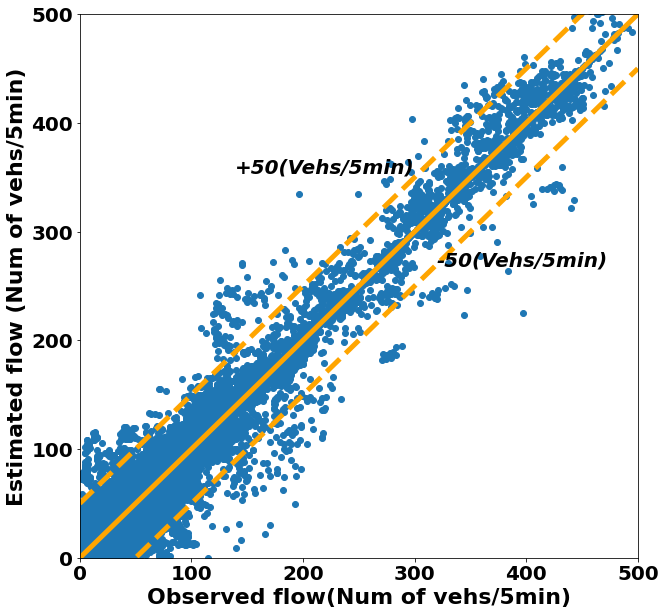}
        \caption{Scatter plot of observed and estimated flows}
        \label{fig:mae_scatter}
    \end{minipage}
\end{figure}
Additionally, a detailed recovery performance evaluation of the AOR framework is visually represented in Figure \ref{fig:mae_dist} and \ref{fig:mae_scatter}. Using PM peak data as an example, Figure \ref{fig:mae_dist} illustrates a histogram showing that approximately 90\% of the links recorded a Mean Absolute Error (MAE) within the desirable range of under 20 vehicles per 5-minute interval. This indicates a high level of accuracy in the estimated traffic flows. Furthermore, the scatter plot in Figure \ref{fig:mae_scatter} corroborates the robustness of the AOR methodology. The plot demonstrates that the majority of estimated flows are closely clustered around the line of good estimation for both lower-demand and higher-demand areas, with deviations largely within a margin of ±50 vehicles per 5-minute interval.

These detailed assessments, visualized through the histogram and scatter plot, underscore the effectiveness of the AOR framework in delivering accurate traffic flow estimations within a macroscopic urban traffic network setting.

%\begin{figure*}[h!]
%\centering
%\begin{subfigure}[a]{\textwidth}
%\centering
%\includegraphics[width=0.45\textwidt%h]{pics/MAE_dist.png}
%\end{subfigure}
%\hfill
%\begin{subfigure}[b]{\textwidth}
%\centering
%\includegraphics[width=0.45\textwidt%h]{pics/MAE_scatter.png}
%\end{subfigure}
%\caption{(a) Histogram plot of the %distribution of links by MAE ranges %and (b) scatter plot of observed %and estimated flows}
%\label{fig:hist_sca}
%\end{figure*}

\begin{figure}[ht]
    \centering
    \includegraphics[width=0.8 \textwidth]{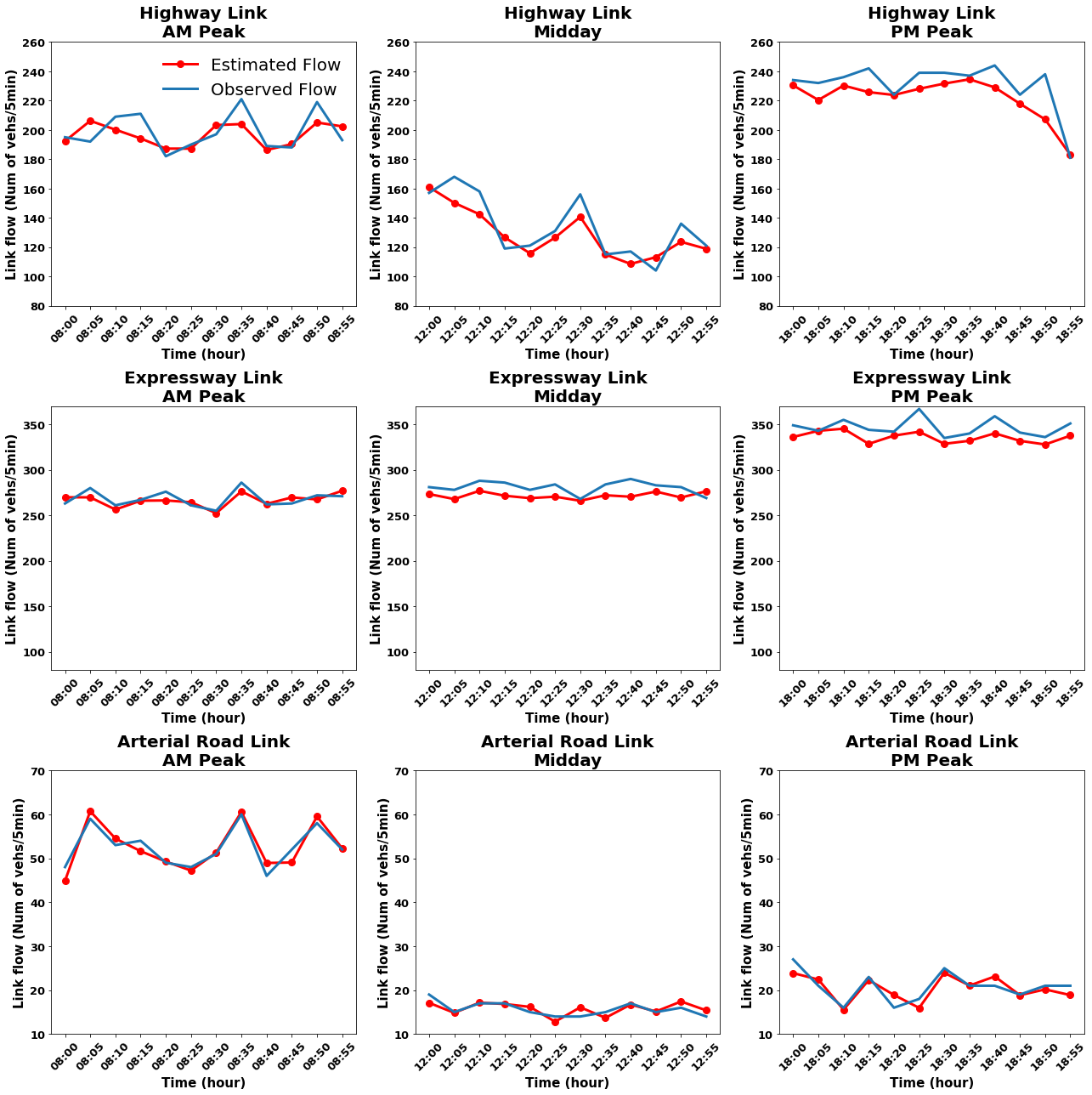}
    \caption{Comparative analysis of average link flow estimation on high-demand road types across different peak periods}
    \label{fig:ana_avg_high}
\end{figure}
The performance of traffic flow recovery is further explored by examining individual links across various road types within a large-scale traffic network at the microscopic level. Leveraging the AOR framework, this study extends the methodologies by selecting links with average traffic flow in each road type for evaluation. Figure \ref{fig:ana_avg_high} presents a comparative analysis of links with average flow estimation results on high-demand road types across different peak periods. It is clear that the estimated flows closely follow the observed flows, with minimal discrepancies noted, even as the observed flow values vary across different times. This alignment indicates that the AOR framework performs robustly in high-demand areas, accurately capturing the variations in traffic flow during peak times. The ability to maintain precision in these areas underscores the model's effectiveness in handling heavy traffic conditions, ensuring reliable traffic management and planning.

\begin{figure}[ht]
    \centering    \includegraphics[width=0.8\textwidth]{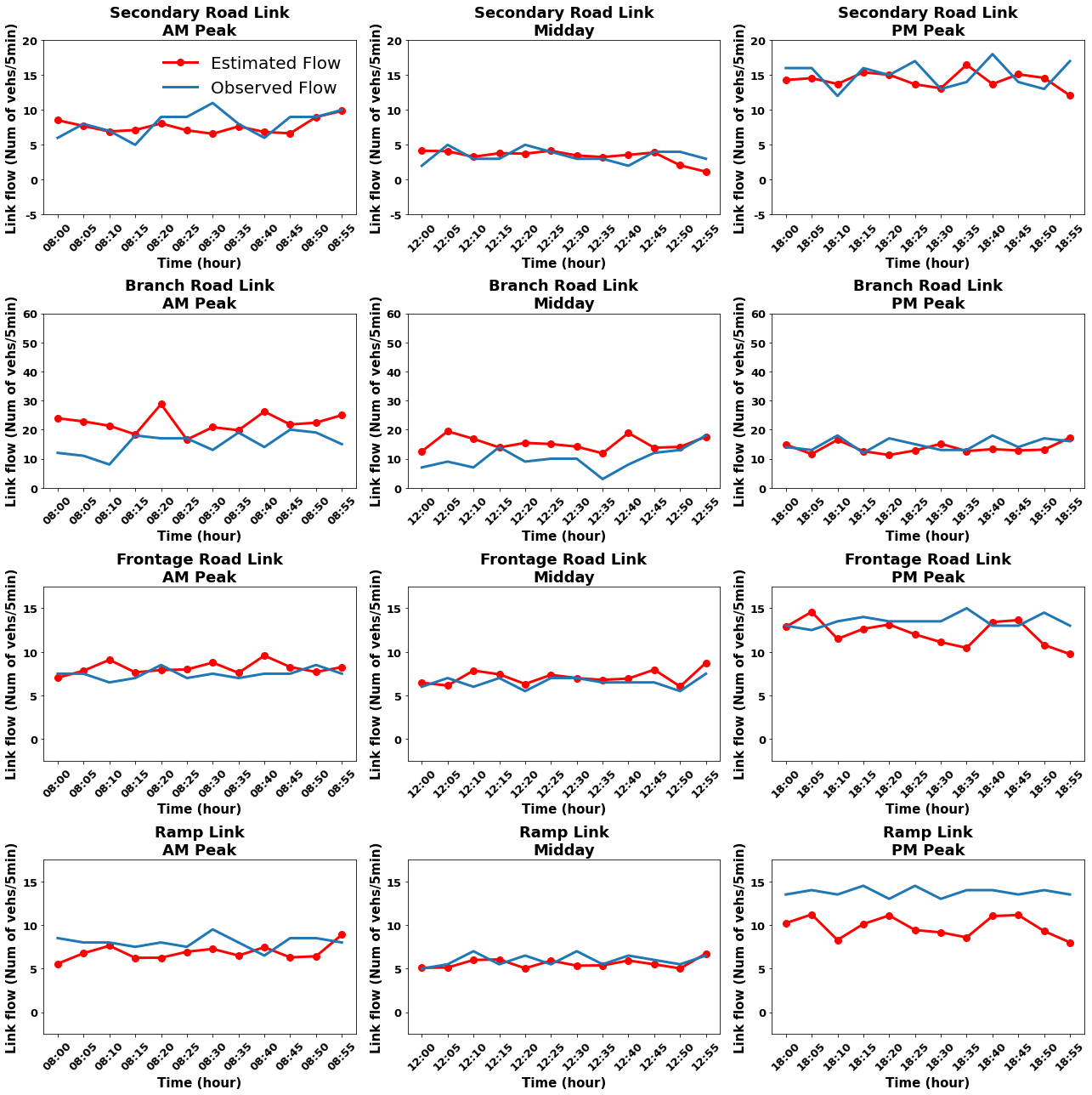}
    \caption{Comparative analysis of average link flow estimation on low-demand road types across different peak periods}
    \label{fig:ana_avg_low}
\end{figure}

Conversely, Figure \ref{fig:ana_avg_low} illustrates the recovery performance for selected links on low-demand road types across different peak periods. Despite occasional discrepancies due to lower traffic volumes, the errors remain within a reasonable range, and most estimated flows closely track the observed flows across various time periods. The AOR framework effectively captures the flow dynamics in these areas, ensuring that even less congested parts of the network are accurately modeled. This comprehensive performance underscores the framework's versatility and robustness in handling diverse traffic scenarios, particularly benefiting complex urban network traffic research.

\subsection{Real World Case}
\begin{figure}[ht]
    \centering
    \includegraphics[width=0.65\textwidth]{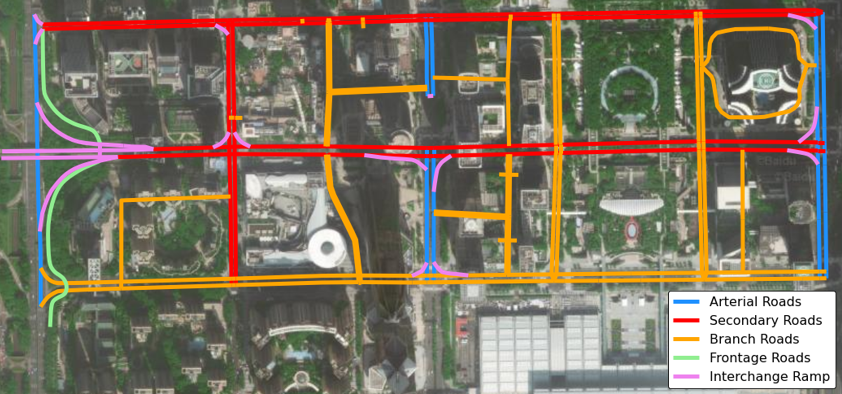}
    \caption{GIS representation of the road network in Futian Central Area}
    \label{fig:GIS_view}
\end{figure}

To demonstrate the effectiveness of the AOR method under real-world conditions, a case study was conducted in the Futian Central Area of Shenzhen, Guangdong Province, China. Given the challenges associated with collecting large-scale urban traffic data, the realistic study focused on a smaller road network within this area. The Futian Central Area spans approximately 1 square kilometer and comprises 173 road junctions and 262 road segments. The GIS representation of this network is shown in Figure \ref{fig:GIS_view}.

Unlike the hypothetical case study using simulation data, this real-world case study utilized actual limited traffic flow and speed data collected from July 5 to July 11, 2021. The data was sourced from GPS-equipped floating cars, mobile devices, and sparsely positioned surveillance cameras within the Futian Central Area. 

Following the previous hypothetical case discussion on evaluation metrics and parameter configurations, the experimental results spanning July 5th to July 11th are systematically articulated in Table \ref{tab:real_res}. This table showcases the performance estimates, including the average WRME, alongside the average solving time for the flow recovery task on each daily dataset.

\begin{table}[ht]
\caption{Recovery performance evaluation of real-world case study}
\centering
\begin{tabular}{l l l l l l l l l }
\hline \textbf{Metrics} & \textbf{July/5}  & \textbf{July/6}  & \textbf{July/7} & \textbf{July/8} & \textbf{July/9} & \textbf{July/10} & \textbf{July/11} & \textbf{Avg}\\
\hline 
Avg solving time & 23.4s & 21.5s & 24.6s & 23.2s & 21.3 & 22.7 & 22.5 &  \textbf{22.7}  \\
\hline Avg WRME & 0.1825 & 0.2087 & 0.1926 & 0.1852 & 0.1904 & 0.1806 & 0.1812 & \textbf{0.1887}  \\
\hline
\end{tabular}
\label{tab:real_res}
\end{table} 

\begin{figure}[ht]
\centering
\includegraphics[width=1 \textwidth]{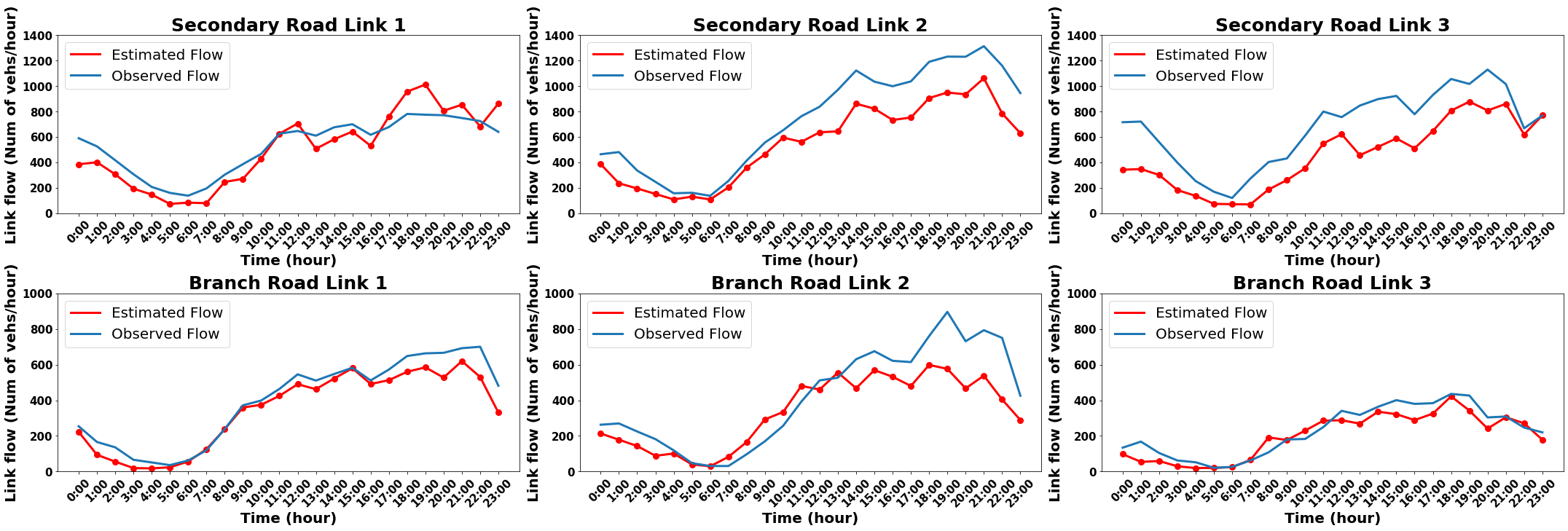}
\caption{Comparison of estimated and observed traffic flow on selected roads in real-world case study}
\label{fig:real_link_estimation}
\end{figure}

The experimental analysis revealed that the average WRME achieved was 0.1887 and the average computational time required was approximately 22.7 seconds. For a more granular examination of recovery performance, the date of July 10th was selected, corresponding to the observed minimum WRME. Specifically, three secondary roads and three branch roads were chosen for a detailed presentation of the estimation results. To enhance visualization, the recovery data, initially calculated at five-minute intervals, was aggregated into hourly metrics for display. The resulting comparison is depicted in Figure \ref{fig:real_link_estimation}, where the red line signifies the estimated traffic flow and the blue line represents the observed flow, illustrating the efficacy of the flow recovery model in approximating real-world traffic conditions.

In the small-scale real-world network case study, the proposed AOR model demonstrates a strong ability to recover network-wide traffic flow values, even with limited observed flow data. The model successfully captures variations in traffic demand throughout the day, maintaining estimation errors within a reasonable range. However, challenges related to the quality and completeness of the collected data impact the model's accuracy. Specifically, the GPS-derived speed data, with nearly 20\% missing coverage in the small network, does not fully capture the progression of traffic flow, which can lead to discrepancies with actual traffic patterns. Additionally, the limited availability of ground-truth flow data across the network hampers a comprehensive evaluation of the model's performance, potentially causing overfitting to areas with available flow data and introducing bias in the estimations. This underscores the importance of the simulation settings discussed in the hypothetical case study, which offer valuable insights for urban traffic flow recovery research. Despite these challenges, the AOR framework shows significant potential for broader application in large-scale urban transportation networks, particularly when GPS link-level speed data is available, making it a promising tool for traffic flow recovery tasks.

\section{Conclusion}
This paper proposes a novel Analytical Optimized Recovery framework to address the challenges of recovering traffic flow from sparse observational data, particularly due to limited traffic sensor coverage. The framework formulates a linear optimization problem to effectively recover network-wide traffic flow patterns under these conditions.

The AOR framework employs a quadratic objective function with 
$l_2$ norm regularization terms to ensure accurate and robust traffic flow recovery. It leverages a dynamic traffic assignment matrix to capture the relationship between link flows and OD demands over discrete time intervals, effectively modeling traffic propagation through the network using GPS speed data. This matrix, combined with limited traffic flow observations, forms a comprehensive system of equations representing traffic dynamics across large-scale urban networks. Lagrangian Relaxation maintains non-negativity constraints, while Stochastic Gradient Descent optimizes hyperparameters, enhancing overall model performance.

Validated through hypothetical case studies in Shenzhen’s Futian District using the SUMO platform, the AOR framework demonstrated the ability to recover dynamic 5-minute link flows with a weighted relative mean error of around 16\%. The experiments included multiple road types within urban traffic networks, showing the framework's robustness in accurately modeling traffic flow dynamics across various road types, from highways to branch roads. Despite the complexity of large-scale urban networks, the AOR framework achieves reasonable solving times and maintains computational efficiency through iterative optimization techniques. This recovered link flow information aids transportation planners and traffic management agencies in understanding the characteristics of dynamic flow and its evolution at the city scale, leading to more informed decision-making in traffic management and planning.

Despite the framework’s effectiveness in recovering traffic flow data, certain limitations must be acknowledged to guide future enhancements. The experimental setup requires additional validation of the observed link locations to ensure robust traffic flow estimates. Moreover, the methodology’s ability to handle conditions with sparse speed data also needs further exploration. Addressing these limitations is crucial for expanding the framework's application, particularly for predictive traffic management and real-time decision support in complex urban traffic networks.

\bibliographystyle{plainnat}
\bibliography{bibliography/biblio}

\begin{thebibliography}{61}
\providecommand{\natexlab}[1]{#1}
\providecommand{\url}[1]{\texttt{#1}}
\expandafter\ifx\csname urlstyle\endcsname\relax
  \providecommand{\doi}[1]{doi: #1}\else
  \providecommand{\doi}{doi: \begingroup \urlstyle{rm}\Url}\fi

\bibitem[Aasen et~al.(2018)Aasen, Honkavaara, Lucieer, and Zarco-Tejada]{aasen2018quantitative}
Helge Aasen, Eija Honkavaara, Arko Lucieer, and Pablo~J Zarco-Tejada.
\newblock Quantitative remote sensing at ultra-high resolution with uav spectroscopy: a review of sensor technology, measurement procedures, and data correction workflows.
\newblock \emph{Remote Sensing}, 10\penalty0 (7):\penalty0 1091, 2018.

\bibitem[Bas et~al.(2007)Bas, Tekalp, and Salman]{bas2007automatic}
Erhan Bas, A~Murat Tekalp, and F~Sibel Salman.
\newblock Automatic vehicle counting from video for traffic flow analysis.
\newblock In \emph{2007 IEEE intelligent vehicles symposium}, pages 392--397. Ieee, 2007.

\bibitem[Bell et~al.(1997)Bell, Shield, Busch, and Kruse]{bell1997stochastic}
Michael~GH Bell, Caroline~M Shield, Fritz Busch, and Gunter Kruse.
\newblock A stochastic user equilibrium path flow estimator.
\newblock \emph{Transportation Research Part C: Emerging Technologies}, 5\penalty0 (3-4):\penalty0 197--210, 1997.

\bibitem[Beymer et~al.(1997)Beymer, McLauchlan, Coifman, and Malik]{beymer1997real}
David Beymer, Philip McLauchlan, Benjamin Coifman, and Jitendra Malik.
\newblock A real-time computer vision system for measuring traffic parameters.
\newblock In \emph{Proceedings of IEEE computer society conference on computer vision and pattern recognition}, pages 495--501. IEEE, 1997.

\bibitem[Bibri(2021)]{bibri2021data}
Simon~Elias Bibri.
\newblock Data-driven smart sustainable cities of the future: Urban computing and intelligence for strategic, short-term, and joined-up planning.
\newblock \emph{Computational Urban Science}, 1\penalty0 (1):\penalty0 8, 2021.

\bibitem[Buch et~al.(2011)Buch, Velastin, and Orwell]{buch2011review}
Norbert Buch, Sergio~A Velastin, and James Orwell.
\newblock A review of computer vision techniques for the analysis of urban traffic.
\newblock \emph{IEEE Transactions on intelligent transportation systems}, 12\penalty0 (3):\penalty0 920--939, 2011.

\bibitem[Chao and Chen(2014)]{chao2014intelligent}
Kuei-Hsiang Chao and Pi-Yun Chen.
\newblock An intelligent traffic flow control system based on radio frequency identification and wireless sensor networks.
\newblock \emph{International journal of distributed sensor networks}, 10\penalty0 (5):\penalty0 694545, 2014.

\bibitem[Coifman et~al.(1998)Coifman, Beymer, McLauchlan, and Malik]{coifman1998real}
Benjamin Coifman, David Beymer, Philip McLauchlan, and Jitendra Malik.
\newblock A real-time computer vision system for vehicle tracking and traffic surveillance.
\newblock \emph{Transportation Research Part C: Emerging Technologies}, 6\penalty0 (4):\penalty0 271--288, 1998.

\bibitem[Daganzo(1977)]{daganzo1977some}
Carlos~F Daganzo.
\newblock Some statistical problems in connection with traffic assignment.
\newblock \emph{Transportation Research}, 11\penalty0 (6):\penalty0 385--389, 1977.

\bibitem[Daganzo and Sheffi(1977)]{daganzo1977stochastic}
Carlos~F Daganzo and Yosef Sheffi.
\newblock On stochastic models of traffic assignment.
\newblock \emph{Transportation science}, 11\penalty0 (3):\penalty0 253--274, 1977.

\bibitem[Daskin(1985)]{daskin1985urban}
Mark~S Daskin.
\newblock Urban transportation networks: Equilibrium analysis with mathematical programming methods, 1985.

\bibitem[Dharia and Adeli(2003)]{dharia2003neural}
Abhijit Dharia and Hojjat Adeli.
\newblock Neural network model for rapid forecasting of freeway link travel time.
\newblock \emph{Engineering Applications of Artificial Intelligence}, 16\penalty0 (7-8):\penalty0 607--613, 2003.

\bibitem[Di~Gangi et~al.(2016)Di~Gangi, Cantarella, Di~Pace, and Memoli]{di2016network}
Massimo Di~Gangi, Giulio~E Cantarella, Roberta Di~Pace, and Silvio Memoli.
\newblock Network traffic control based on a mesoscopic dynamic flow model.
\newblock \emph{Transportation Research Part C: Emerging Technologies}, 66:\penalty0 3--26, 2016.

\bibitem[Dial(1971)]{dial1971probabilistic}
Robert~B Dial.
\newblock A probabilistic multipath traffic assignment model which obviates path enumeration.
\newblock \emph{Transportation research}, 5\penalty0 (2):\penalty0 83--111, 1971.

\bibitem[Essien et~al.(2021)Essien, Petrounias, Sampaio, and Sampaio]{essien2021deep}
Aniekan Essien, Ilias Petrounias, Pedro Sampaio, and Sandra Sampaio.
\newblock A deep-learning model for urban traffic flow prediction with traffic events mined from twitter.
\newblock \emph{World Wide Web}, 24\penalty0 (4):\penalty0 1345--1368, 2021.

\bibitem[Fabris et~al.(2024)Fabris, Ceccato, and Zanella]{fabris2024efficient}
Marco Fabris, Riccardo Ceccato, and Andrea Zanella.
\newblock Efficient sensors selection for traffic flow monitoring: An overview of model-based techniques leveraging network observability.
\newblock \emph{arXiv preprint arXiv:2404.08588}, 2024.

\bibitem[Fan et~al.(2020)Fan, Wong, Zhang, and Du]{fan2020dynamically}
Tianxiang Fan, SC~Wong, Zhiwen Zhang, and Jie Du.
\newblock A dynamically bi-orthogonal solution method for a stochastic lighthill-whitham-richards traffic flow model.
\newblock \emph{Computer-Aided Civil and Infrastructure Engineering}, 2020.

\bibitem[Fedorov et~al.(2019)Fedorov, Nikolskaia, Ivanov, Shepelev, and Minbaleev]{fedorov2019traffic}
Aleksandr Fedorov, Kseniia Nikolskaia, Sergey Ivanov, Vladimir Shepelev, and Alexey Minbaleev.
\newblock Traffic flow estimation with data from a video surveillance camera.
\newblock \emph{Journal of Big Data}, 6:\penalty0 1--15, 2019.

\bibitem[Fisk(1977)]{fisk1977note}
C~Fisk.
\newblock Note on the maximum likelihood calibration on dial's assignment method.
\newblock \emph{Transportation Research}, 11\penalty0 (1):\penalty0 67--68, 1977.

\bibitem[Fisk(1980)]{fisk1980some}
Caroline Fisk.
\newblock Some developments in equilibrium traffic assignment.
\newblock \emph{Transportation Research Part B: Methodological}, 14\penalty0 (3):\penalty0 243--255, 1980.

\bibitem[Florian and Chen(1995)]{florian1995coordinate}
Michael Florian and Yang Chen.
\newblock A coordinate descent method for the bi-level od matrix adjustment problem.
\newblock \emph{International Transactions in Operational Research}, 2\penalty0 (2):\penalty0 165--179, 1995.

\bibitem[Id{\'e} et~al.(2016)Id{\'e}, Katsuki, Morimura, and Morris]{ide2016city}
Tsuyoshi Id{\'e}, Takayuki Katsuki, Tetsuro Morimura, and Robert Morris.
\newblock City-wide traffic flow estimation from a limited number of low-quality cameras.
\newblock \emph{IEEE Transactions on Intelligent Transportation Systems}, 18\penalty0 (4):\penalty0 950--959, 2016.

\bibitem[Iordanidou et~al.(2014)Iordanidou, Papamichail, Roncoli, and Papageorgiou]{iordanidou2014feedback}
Georgia-Roumpini Iordanidou, Ioannis Papamichail, Claudio Roncoli, and Markos Papageorgiou.
\newblock A feedback-based approach for mainstream traffic flow control of multiple bottlenecks on motorways.
\newblock \emph{IFAC Proceedings Volumes}, 47\penalty0 (3):\penalty0 11344--11349, 2014.

\bibitem[Jiang and Adeli(2004)]{jiang2004wavelet}
Xiaomo Jiang and Hojjat Adeli.
\newblock Wavelet packet-autocorrelation function method for traffic flow pattern analysis.
\newblock \emph{Computer-Aided Civil and Infrastructure Engineering}, 19\penalty0 (5):\penalty0 324--337, 2004.

\bibitem[Jiang and Adeli(2005)]{jiang2005dynamic}
Xiaomo Jiang and Hojjat Adeli.
\newblock Dynamic wavelet neural network model for traffic flow forecasting.
\newblock \emph{Journal of transportation engineering}, 131\penalty0 (10):\penalty0 771--779, 2005.

\bibitem[Kashyap et~al.(2022)Kashyap, Raviraj, Devarakonda, Nayak~K, KV, and Bhat]{kashyap2022traffic}
Anirudh~Ameya Kashyap, Shravan Raviraj, Ananya Devarakonda, Shamanth~R Nayak~K, Santhosh KV, and Soumya~J Bhat.
\newblock Traffic flow prediction models--a review of deep learning techniques.
\newblock \emph{Cogent Engineering}, 9\penalty0 (1):\penalty0 2010510, 2022.

\bibitem[Khan et~al.(2017)Khan, Sargento, and Luis]{khan2017data}
Muhammad~Awais Khan, Susana Sargento, and Miguel Luis.
\newblock Data collection from smart-city sensors through large-scale urban vehicular networks.
\newblock In \emph{2017 IEEE 86th Vehicular Technology Conference (VTC-Fall)}, pages 1--6. IEEE, 2017.

\bibitem[Khoshabeh et~al.(2007)Khoshabeh, Gandhi, and Trivedi]{khoshabeh2007multi}
Ramsin Khoshabeh, Tarak Gandhi, and Mohan~M Trivedi.
\newblock Multi-camera based traffic flow characterization \& classification.
\newblock In \emph{2007 IEEE Intelligent Transportation Systems Conference}, pages 259--264. IEEE, 2007.

\bibitem[LeBlanc and Farhangian(1982)]{leblanc1982selection}
Larry~J LeBlanc and Keyvan Farhangian.
\newblock Selection of a trip table which reproduces observed link flows.
\newblock \emph{Transportation Research Part B: Methodological}, 16\penalty0 (2):\penalty0 83--88, 1982.

\bibitem[Lei et~al.(2022)Lei, Mei, Shi, and Wei]{lei2022modeling}
Xiaoliang Lei, Hao Mei, Bin Shi, and Hua Wei.
\newblock Modeling network-level traffic flow transitions on sparse data.
\newblock In \emph{Proceedings of the 28th ACM SIGKDD Conference on Knowledge Discovery and Data Mining}, pages 835--845, 2022.

\bibitem[Li and Chen(2023)]{li2023strategy}
Guoyuan Li and Anthony Chen.
\newblock Strategy-based transit stochastic user equilibrium model with capacity and number-of-transfers constraints.
\newblock \emph{European Journal of Operational Research}, 305\penalty0 (1):\penalty0 164--183, 2023.

\bibitem[Li et~al.(2021)Li, Boonaert, Doniec, and Lozenguez]{li2021multi}
Jinjian Li, Jacques Boonaert, Arnaud Doniec, and Guillaume Lozenguez.
\newblock Multi-models machine learning methods for traffic flow estimation from floating car data.
\newblock \emph{Transportation Research Part C: Emerging Technologies}, 132:\penalty0 103389, 2021.

\bibitem[Li et~al.(2022{\natexlab{a}})Li, Wu, Li, Pian, Huang, Xu, and Li]{li2022st}
Jinlong Li, Pan Wu, Ruonan Li, Yuzhuang Pian, Zilin Huang, Lunhui Xu, and Xiaochen Li.
\newblock St-crmf: compensated residual matrix factorization with spatial-temporal regularization for graph-based time series forecasting.
\newblock \emph{Sensors}, 22\penalty0 (15):\penalty0 5877, 2022{\natexlab{a}}.

\bibitem[Li et~al.(2018)Li, Min, Jia-qing, Ling-yu, Ke, and Zheng-xi]{li2018urban}
WANG Li, LI~Min, YAN Jia-qing, ZHANG Ling-yu, PAN Ke, and LI~Zheng-xi.
\newblock Urban traffic flow data recovery method based on generative adversarial network.
\newblock \emph{Journal of Transportation Systems Engineering and Information Technology}, 18\penalty0 (6):\penalty0 63, 2018.

\bibitem[Li et~al.(2022{\natexlab{b}})Li, Yang, and Jabari]{li2022nonlinear}
Wenqing Li, Chuhan Yang, and Saif~Eddin Jabari.
\newblock Nonlinear traffic prediction as a matrix completion problem with ensemble learning.
\newblock \emph{Transportation science}, 56\penalty0 (1):\penalty0 52--78, 2022{\natexlab{b}}.

\bibitem[Lin et~al.(2021)Lin, Zhang, He, Feng, Wu, and Li]{lin2021vehicle}
Zongyu Lin, Guozhen Zhang, Zhiqun He, Jie Feng, Wei Wu, and Yong Li.
\newblock Vehicle trajectory recovery on road network based on traffic camera video data.
\newblock In \emph{Proceedings of the 29th International Conference on Advances in Geographic Information Systems}, pages 389--398, 2021.

\bibitem[Liu et~al.(2021{\natexlab{a}})Liu, Shi, Kiani, Khreishah, Lee, Ansari, Liu, and Yousef]{liu2021smart}
Guanxiong Liu, Hang Shi, Abbas Kiani, Abdallah Khreishah, Joyoung Lee, Nirwan Ansari, Chengjun Liu, and Mustafa~Mohammad Yousef.
\newblock Smart traffic monitoring system using computer vision and edge computing.
\newblock \emph{IEEE Transactions on Intelligent Transportation Systems}, 23\penalty0 (8):\penalty0 12027--12038, 2021{\natexlab{a}}.

\bibitem[Liu and Liu(2023)]{liu2023stochastic}
Ke~Liu and Yanli Liu.
\newblock Stochastic user equilibrium based spatial-temporal distribution prediction of electric vehicle charging load.
\newblock \emph{Applied Energy}, 339:\penalty0 120943, 2023.

\bibitem[Liu et~al.(2021{\natexlab{b}})Liu, Lyu, Zhang, Liu, Yu, and Qu]{liu2021deeptsp}
Yang Liu, Cheng Lyu, Yuan Zhang, Zhiyuan Liu, Wenwu Yu, and Xiaobo Qu.
\newblock Deeptsp: Deep traffic state prediction model based on large-scale empirical data.
\newblock \emph{Communications in transportation research}, 1:\penalty0 100012, 2021{\natexlab{b}}.

\bibitem[Lu et~al.(2013)Lu, Zhou, and Zhang]{lu2013dynamic}
Chung-Cheng Lu, Xuesong Zhou, and Kuilin Zhang.
\newblock Dynamic origin--destination demand flow estimation under congested traffic conditions.
\newblock \emph{Transportation Research Part C: Emerging Technologies}, 34:\penalty0 16--37, 2013.

\bibitem[Ma and Qian(2018)]{ma2018estimating}
Wei Ma and Zhen~Sean Qian.
\newblock Estimating multi-year 24/7 origin-destination demand using high-granular multi-source traffic data.
\newblock \emph{Transportation Research Part C: Emerging Technologies}, 96:\penalty0 96--121, 2018.

\bibitem[Ma et~al.(2020)Ma, Pi, and Qian]{ma2020estimating}
Wei Ma, Xidong Pi, and Sean Qian.
\newblock Estimating multi-class dynamic origin-destination demand through a forward-backward algorithm on computational graphs.
\newblock \emph{Transportation Research Part C: Emerging Technologies}, 119:\penalty0 102747, 2020.

\bibitem[Manguri and Mohammed(2023)]{manguri2023review}
Kamaran Hussein~Khdir Manguri and Aree~Ali Mohammed.
\newblock A review of computer vision--based traffic controlling and monitoring.
\newblock \emph{UHD Journal of Science and Technology}, 7\penalty0 (2):\penalty0 6--15, 2023.

\bibitem[Meng et~al.(2017)Meng, Yi, Su, Gao, and Zheng]{meng2017city}
Chuishi Meng, Xiuwen Yi, Lu~Su, Jing Gao, and Yu~Zheng.
\newblock City-wide traffic volume inference with loop detector data and taxi trajectories.
\newblock In \emph{Proceedings of the 25th ACM SIGSPATIAL International Conference on Advances in Geographic Information Systems}, pages 1--10, 2017.

\bibitem[Necula(2014)]{necula2014dynamic}
Emilian Necula.
\newblock Dynamic traffic flow prediction based on gps data.
\newblock In \emph{2014 IEEE 26th International Conference on Tools with Artificial Intelligence}, pages 922--929. IEEE, 2014.

\bibitem[Nguyen(1977)]{nguyen1977estimating}
S~Nguyen.
\newblock Estimating an od matrix from network data: A network equilibrium approach, publication 87.
\newblock \emph{Centre de Recherche sur les Transports, Universite de Montreal}, 1977.

\bibitem[Qi et~al.(2019)Qi, Zhao, Zhang, Jin, Wang, and Runge]{qi2019automated}
Bozhao Qi, Wei Zhao, Haiping Zhang, Zhihong Jin, Xiaohan Wang, and Troy Runge.
\newblock Automated traffic volume analytics at road intersections using computer vision techniques.
\newblock In \emph{2019 5th International Conference on Transportation Information and Safety (ICTIS)}, pages 161--169. IEEE, 2019.

\bibitem[Ran et~al.(2016)Ran, Tan, Wu, and Jin]{ran2016tensor}
Bin Ran, Huachun Tan, Yuankai Wu, and Peter~J Jin.
\newblock Tensor based missing traffic data completion with spatial--temporal correlation.
\newblock \emph{Physica A: Statistical Mechanics and its Applications}, 446:\penalty0 54--63, 2016.

\bibitem[Tang and Zeng(2022)]{tang2022spatiotemporal}
Jinjun Tang and Jie Zeng.
\newblock Spatiotemporal gated graph attention network for urban traffic flow prediction based on license plate recognition data.
\newblock \emph{Computer-Aided Civil and Infrastructure Engineering}, 37\penalty0 (1):\penalty0 3--23, 2022.

\bibitem[Unterluggauer et~al.(2022)Unterluggauer, Rich, Andersen, and Hashemi]{unterluggauer2022electric}
Tim Unterluggauer, Jeppe Rich, Peter~Bach Andersen, and Seyedmostafa Hashemi.
\newblock Electric vehicle charging infrastructure planning for integrated transportation and power distribution networks: A review.
\newblock \emph{ETransportation}, 12:\penalty0 100163, 2022.

\bibitem[Wardrop(1952)]{wardrop1952road}
John~Glen Wardrop.
\newblock Road paper. some theoretical aspects of road traffic research.
\newblock \emph{Proceedings of the institution of civil engineers}, 1\penalty0 (3):\penalty0 325--362, 1952.

\bibitem[Wei et~al.(2019)Wei, Shi, Yang, Qian, Ji, and Jiang]{wei2019city}
Peter Wei, Haocong Shi, Jiaying Yang, Jingyi Qian, Yinan Ji, and Xiaofan Jiang.
\newblock City-scale vehicle tracking and traffic flow estimation using low frame-rate traffic cameras.
\newblock In \emph{Adjunct Proceedings of the 2019 ACM International Joint Conference on Pervasive and Ubiquitous Computing and Proceedings of the 2019 ACM International Symposium on Wearable Computers}, pages 602--610, 2019.

\bibitem[Xia et~al.(2016)Xia, Shi, Song, Geng, and Liu]{xia2016towards}
Yingjie Xia, Xingmin Shi, Guanghua Song, Qiaolei Geng, and Yuncai Liu.
\newblock Towards improving quality of video-based vehicle counting method for traffic flow estimation.
\newblock \emph{Signal Processing}, 120:\penalty0 672--681, 2016.

\bibitem[Yang et~al.(1992)Yang, Sasaki, Iida, and Asakura]{yang1992estimation}
Hai Yang, Tsuna Sasaki, Yasunori Iida, and Yasuo Asakura.
\newblock Estimation of origin-destination matrices from link traffic counts on congested networks.
\newblock \emph{Transportation Research Part B: Methodological}, 26\penalty0 (6):\penalty0 417--434, 1992.

\bibitem[Yang et~al.(2001)Yang, Meng, and Bell]{yang2001simultaneous}
Hai Yang, Qiang Meng, and Michael~GH Bell.
\newblock Simultaneous estimation of the origin-destination matrices and travel-cost coefficient for congested networks in a stochastic user equilibrium.
\newblock \emph{Transportation science}, 35\penalty0 (2):\penalty0 107--123, 2001.

\bibitem[Yi et~al.(2019)Yi, Duan, Li, Li, Zhang, and Zheng]{yi2019citytraffic}
Xiuwen Yi, Zhewen Duan, Ting Li, Tianrui Li, Junbo Zhang, and Yu~Zheng.
\newblock Citytraffic: Modeling citywide traffic via neural memorization and generalization approach.
\newblock In \emph{Proceedings of the 28th ACM international conference on information and knowledge management}, pages 2665--2671, 2019.

\bibitem[Yu et~al.(2020)Yu, Stettler, Angeloudis, Hu, and Chen]{yu2020urban}
Jingru Yu, Marc~EJ Stettler, Panagiotis Angeloudis, Simon Hu, and Xiqun~Michael Chen.
\newblock Urban network-wide traffic speed estimation with massive ride-sourcing gps traces.
\newblock \emph{Transportation Research Part C: Emerging Technologies}, 112:\penalty0 136--152, 2020.

\bibitem[Zhang et~al.(2003)Zhang, Wang, Nihan, and Hallenbeck]{zhang2003development}
Xiaoping Zhang, Yinhai Wang, Nancy~L Nihan, and Mark~E Hallenbeck.
\newblock Development of a system for collecting loop-detector event data for individual vehicles.
\newblock \emph{Transportation research record}, 1855\penalty0 (1):\penalty0 168--175, 2003.

\bibitem[Zhang et~al.(2020)Zhang, Li, Lin, and Wang]{zhang2020network}
Zhengchao Zhang, Meng Li, Xi~Lin, and Yinhai Wang.
\newblock Network-wide traffic flow estimation with insufficient volume detection and crowdsourcing data.
\newblock \emph{Transportation Research Part C: Emerging Technologies}, 121:\penalty0 102870, 2020.

\bibitem[Zhou et~al.(2013)Zhou, Wang, and Yu]{zhou2013traffic}
Xiangyu Zhou, Wenjun Wang, and Long Yu.
\newblock Traffic flow analysis and prediction based on gps data of floating cars.
\newblock In \emph{Proceedings of the 2012 International Conference on Information Technology and Software Engineering: Information Technology}, pages 497--508. Springer, 2013.

\bibitem[Zhu et~al.(2022)Zhu, Wu, and Xiao]{zhu2022research}
Yuyu Zhu, QingE Wu, and Na~Xiao.
\newblock Research on highway traffic flow prediction model and decision-making method.
\newblock \emph{Scientific reports}, 12\penalty0 (1):\penalty0 19919, 2022.

\end{thebibliography}

\end{document}